\documentclass[fleqn,usenatbib]{mnras}

\usepackage{comment}
\usepackage{newtxtext,newtxmath}

\usepackage[T1]{fontenc}

\DeclareRobustCommand{\VAN}[3]{#2}
\let\VANthebibliography\thebibliography
\def\thebibliography{\DeclareRobustCommand{\VAN}[3]{##3}\VANthebibliography}

\usepackage{graphicx}	
\usepackage{amsmath}	
\usepackage{amssymb}	

\usepackage{epstopdf}
\usepackage{physics}
\usepackage{float}
\usepackage{enumitem}

\newcommand{\mps}{m s$^{-1}$}
\newcommand{\tess}{\textit{TESS}}
\newcommand{\me}{$M_{\oplus}$}
\newcommand{\re}{$R_{\oplus}$}
\newcommand{\gcm}{g cm$^{-3}$}

\newcommand{\ms}{\ensuremath{m\,s^{-1}}}

\newcommand{\juliet}{\texttt{juliet}}

\title[A 55-day period dense Neptune orbiting around HD 95338]{The Magellan/PFS Exoplanet Search: A 55-day period dense Neptune transiting the bright ($V=8.6$) star HD 95338$^{\dagger \ddagger}$}
\author[M. R. D\'iaz et al.]{
Mat\'ias R. D\'iaz,$^{1,\star}$ 
James S. Jenkins,$^{1}$
Fabo Feng,$^{2}$
R. Paul Butler,$^{2}$
Mikko Tuomi,$^{3}$
\newauthor
~Stephen A. Shectman,$^{4}$
Daniel Thorngren,$^{5,6}$
Maritza G. Soto,$^{7}$
Jos\'e I. Vines,$^{1}$
\newauthor
~Johanna K. Teske,$^{4,11}$
Diana Dragomir,$^{8}$
Steven Villanueva,$^{9}$
Stephen R. Kane,$^{10}$
\newauthor
~Zaira M. Berdi\~{n}as,$^{1}$
Jeffrey D. Crane,$^{4}$
Sharon X. Wang,$^{2,4}$
Pamela Arriagada$^{2}$
\\\\
$^{1}$ Departamento de Astronom\'ia, Universidad de Chile, Camino El Observatorio 1515, Las Condes, Santiago, Chile\\
$^{2}$ Earth and Planets Laboratory, Carnegie Institution for Science, 5124 Broad Branch Road, Washington, DC 20015-1305, USA\\
$^{3}$ Center for Astrophysics Research, School of Physics, Astronomy and Mathematics, University of Hertfordshire, College Lane, Hatfield AL109AB, UK\\
$^{4}$ The Observatories, Carnegie Institution for Science, 813 Santa Barbara Street, Pasadena, CA 91101, USA\\
$^{5}$ Department of Physics, University of California, Santa Cruz, USA\\
$^{6}$ Institut de Recherche sur les Exoplan\`etes, Universit\'e de Montr\'eal, Canada\\
$^{7}$School of Physics and Astronomy, Queen Mary University of London, G.O. Jones Building, 327 Mile End Road London, E1 4NS, UK\\
$^{8}$ Department of Physics and Astronomy, University of New Mexico, 1919 Lomas Blvd NE, Albuquerque, NM 87131, USA\\
$^{9}$ MIT Kavli Institute for Astrophysics \& Space Research, 77 Massachussetts Ave. Building 37-582 BB, Cambridge, MA 02139, USA\\
$^{10}$ Department of Earth and Planetary Sciences, Univerisity of California Riverside, 900 University Ave., Riverside, CA 92521, USA\\ 
$^{11}$ NASA Hubble Fellow\\
$\dagger$ This paper includes data gathered with the 6.5 m Magellan Telescopes located at Las Campanas Observatory, Chile.\\
$\ddagger$Additional observations were acquired with the ESO-3.6m telescope at La Silla Observatory under programs 0101.C-0497, 0102.C-0525 and 0103.C-0442.\\
$^{\star}$ corresponding author, \url{matias.diaz.m@ug.uchile.cl}
}

\date{Accepted XXX. Received YYY; in original form YYZ}

\pubyear{2020}

\begin{document}
\label{firstpage}
\pagerange{\pageref{firstpage}--\pageref{lastpage}}
\maketitle

\begin{abstract}
{We report the detection of a transiting, dense Neptune planet candidate orbiting the bright ($V=8.6$) K0.5V star HD 95338. Detection of the 55-day periodic signal comes from the analysis of precision radial velocities from the Planet Finder Spectrograph on the Magellan II Telescope. Follow-up observations with HARPS also confirm the presence of the periodic signal in the combined data. 
HD 95338 was also observed by the Transiting Exoplanet Survey Satellite ({\it TESS}) where we identify a clear single transit in the photometry.  A Markov Chain Monte Carlo period search on the velocities allows strong constraints on the expected transit time, matching well the epoch calculated from \tess{} data, confirming both signals describe the same companion. A joint fit model yields an absolute mass of 42.44$^{+2.22}_{-2.08} M_{\oplus}$ and a radius of 3.89$^{+0.19}_{-0.20}$ $R_{\oplus}$ which translates to a density of 3.98$^{+0.62}_{-0.64}$ \gcm\, for the planet. Given the planet mass and radius, structure models suggest it is composed of a mixture of ammonia, water, and methane. HD 95338\,b is one of the most dense Neptune planets yet detected, indicating a heavy element enrichment of $\sim$90\% ($\sim38\, M_{\oplus}$).  This system presents a unique opportunity for future follow-up observations that can further constrain structure models of cool gas giant planets.}
\end{abstract}




\section{Introduction}
As the transit probability of a planet orbiting a star decreases with increasing orbital period, or star-planet separation, the majority of transiting systems contain planets with orbital periods of less than 10~days. For planets with longer periods, not only does the probability decrease compared with the  shorter period counterparts, but they are also much more difficult to detect and confirm logistically, using ground-based transit surveys.  Large-scale surveys have been setup to try to target longer period transiting systems (e.g., HATSouth, \citealp{Bakos2013}; NGTS, \citealp{Wheatley2017}), but they are generally limited to detection sensitivities that fall off after 12 days, due to the observing window function problem \citep{Bakos2013}. {Space-based surveys can bypass this issue, as they are capable of monitoring these targets almost continuously.}

The CoRoT \citep{Baglin2006}, {\it Kepler} \citep{Borucki2010a}, and {\it K2} \citep{Howell2014} space missions paved the way for the Transiting Exoplanet Survey Satellite (\citealp[{\it TESS};][]{Ricker2015}) mission.  CoRoT, and {\it Kepler} in particular, were able to provide some startling discoveries, particularly giving a first glimpse into the structural properties of small planets (e.g., CoRoT-7b, \citealp{Corot7_Leger2009};  Kepler-10 b, \citealp{Batalha2011}).  However, what we have learned about giant planets has mainly come from ground-based planet detections, due in no small part to the ease of radial-velocity (RV) follow-up that is a requirement to constrain the mass and density of transit detections. 

Detailed studies have been possible for a handful of gas giant planets. For example, two of the most well-known planets are HD 189733\,b \citep{HD189733_Bouchy2005} and HD 209458\,b \citep{HD209458_Henry2000}. HD 209458\,b was the first confirmed transiting planet \citep{Charbonneau2000} and was also the first that allowed us to detect elements in its escaping atmosphere, in this case Na and CO \citep{Charbonneau2002}. HD 189733\,b also orbits a fairly bright star, and therefore we also found this object to have an inflated atmosphere that is in the process of being evaporated due to the close proximity of the host star \citep{Lecavelier2012,Bourrier2013}. From its escaping atmosphere Sodium D absorption has been characterized \citep{Wyttenbach2015,Salz2016}. Recent studies have revealed water vapor absorption on the planet's atmosphere \citep{Birkby2013, AlonsoFiorano2019} and also absorption due to methane \citep{Brogi2018}.  Beyond these two planets, we now have a number of transiting gas giants that have revealed their atmospheric make-up (e.g., GJ 3470\,b, \citealp{Nascimbeni2013}; WASP-12\,b, \citealp{Kreidberg2015Wasp12Water}; MASCARA-2\,b/KELT-20\,b, \citealp{Casasayas-Barris2019}; KELT-9\,b, \citealp{Turner2020}). 

Although we have learned a great deal about gas giants, the vast majority of what we know applies only to the hottest subset, those closest to their stars that are heavily irradiated.  The equilibrium temperatures of these hot Jupiters are generally $>$1000~K, and therefore their atmospheric chemistries and physical properties are very different to those on longer period orbits, like Jupiter in our solar system. The population of longer period transiting planets is growing (e.g. HATS-17 b, \citealp{Brahm2016}; Kepler-538 b, \citealp{Mayo2019}; EPIC 249893012 c \& d, \citealp{Hidalgo2020}), particularly since the introduction of TESS that finds transits orbiting significantly brighter stars than {\it Kepler} or {\it K2}, and across the whole sky (e.g., HD 1397\,b, \citealp{Brahm2019}; TOI-667\,b, \citealp{Jordan2019}; HD 21749\,b \& c, \citealp{Dragomir2019},  LTT 9779 b, \citealp{Jenkins2020}). However, despite these gains, we still know of not many known transiting planets with orbital periods greater than 40~days, orbiting stars bright enough for detailed atmospheric characterization ($V~<~9$).

Here we introduce HD 95338\,b, a super-Neptune planet detected using precision RVs as part of the Planet Finder Spectrograph (PFS; \citealp{Crane2006,Crane2008,Crane2010}) long term planet search project, and which we found to transit after analyzing the \tess\, lightcurve.  HD 95338 b is the first planet candidate from TESS discovered with a period larger than 27 days (the time baseline of the TESS data series). Therefore, it is the first single-transit planet detected from the \tess\, mission.

\section{Spectroscopic Observations}\label{sec:obs}
High-precision Doppler measurements of HD 95338 were acquired using PFS mounted on the 6.5 m Magellan II (Clay) telescope at Las Campanas Observatory, and the High Accuracy Radial velocity Planet Searcher (\citealp[HARPS;][]{Pepe2002}) installed on the ESO 3.6 m telescope at La Silla Observatory.

\subsection{PFS}

Observations were carried out using PFS between February 26 2010 and May 25 2018, as part of the Magellan Exoplanet Long Term Survey (LTS).
PFS uses an iodine cell for precise RV measurements and it delivers a resolving power of $R\sim$80,000 in the iodine region when observing with the 0.5''$\times$2.5'' slit. Iodine-free template observations were acquired with the 0.3''$\times$2.5'' slit at a resolving power of $R\sim$127,000. 52 observations were acquired using an average of 540 s of exposure time yielding a mean radial velocity uncertainty of 1.13 ~\mps and a median SNR$\sim$144.

PFS was upgraded with a new CCD detector in 2017. The new CCD is a 10k$\times$10k sensor and has smaller pixels, which improves the line sampling in the spectra. In addition, regular LTS stars are now observed using the 0.3"$\times$2.5" slit, therefore improving the resolution. The data using this new setup is labeled as PFS2 and includes 31 observations. For this upgraded setup, the mean exposure time used was 485 s for each observation giving rise to a mean radial velocity uncertainty of 0.87 ~\mps for a median SNR$\sim$74. The radial velocities are computed with a custom pipeline following the procedure outlined by \citet{Butler1996}. They are listed in Table \ref{tab:pfs1rvs} and \ref{tab:pfs2rvs}.

The spectral wavelength range in PFS covers the Ca {\sc ii} H \& K lines, enabling the possibility of deriving S-indices to monitor the stellar chromospheric activity. S-indices are derived using the prescription outlined by \citet{Baliunas1996} and \citet{Boisse2011}. In general, authors determine their S-index errors based on photon noise on the CCD \citep{Boisse2011,Lovis2011, Jenkins2017}. In our case, however, doing so can grossly underestimate the real error, reporting $<1$\% or smaller, as they are probably dominated by instrumental systematics (e.g., wavelength calibration, normalization errors). To avoid any bias to unrealistic error estimation we assumed a homogeneous 5\% errorbar estimated from the RMS of the S-index series.

\begin{table*}
\center
\caption{PFS1 Radial Velocities of HD 95338. This table is published in its entirety in the machine-readable format. A portion is shown here for guidance regarding its form and content.}
\label{tab:pfs1rvs}
\begin{tabular}{lcccc}
\hline\hline
\multicolumn{1}{l}{BJD}& \multicolumn{1}{c}{RV}&\multicolumn{1}{c}{$\sigma$ RV}& \multicolumn{1}{c}{S}&\multicolumn{1}{c}{$\sigma$ S}\\ 
\multicolumn{1}{l}{(- 2450000)}& \multicolumn{1}{c}{(m s$^{-1}$)}&\multicolumn{1}{c}{(m s$^{-1}$)}& \multicolumn{1}{c}{(dex)}&\multicolumn{1}{c}{(dex)}\\ \hline 
5253.72066 & 1.806 & 1.191 & 0.2450 & 0.012 \\
5256.80073 & 3.796 & 1.186 & 0.1867 & 0.012 \\
5342.53484 & -2.873 & 1.114 & 0.3596 & 0.012 \\
5348.50146 & 0.620 & 1.317 & 0.2815 & 0.012\\
5349.52059 & -1.081 & 1.371 & 0.2713 & 0.012 \\
5588.85377 & 2.115 & 0.988 & 0.1724 & 0.012 \\
5663.60446 & 5.616 & 1.178 & 0.1918 & 0.012 \\
5959.79501 & -3.994 & 1.019 & 0.2402 & 0.012 \\
6284.83957 & -6.118 & 0.836 & 0.2481 & 0.012 \\
6291.83583 & -7.558 & 0.829 & 0.1590 & 0.012 \\
6345.74970 & -6.404 & 1.179 & 0.2418 & 0.012 \\
6355.71078 & -2.553 & 1.206 & 0.3401 & 0.012 \\
...&...&...&...&...\\\hline
\end{tabular}
\end{table*}

\begin{table*}
\center
\caption{PFS2 Radial Velocities of HD 95338. This table is published in its entirety in the machine-readable format. A portion is shown here for guidance regarding its form and content.}
\label{tab:pfs2rvs}
\begin{tabular}{lcccc}
\hline\hline
\multicolumn{1}{l}{BJD}& \multicolumn{1}{c}{RV}&\multicolumn{1}{c}{$\sigma$ RV}& \multicolumn{1}{c}{S}&\multicolumn{1}{c}{$\sigma$ S}\\ 
\multicolumn{1}{l}{(- 2450000)}& \multicolumn{1}{c}{(m s$^{-1}$)}&\multicolumn{1}{c}{(m s$^{-1}$)}& \multicolumn{1}{c}{(dex)}&\multicolumn{1}{c}{(dex)}\\ \hline 
8471.81505 & 5.205 & 0.931 & 0.1644 & 0.008 \\
8471.82063 & 3.733 & 0.892 & 0.1659 & 0.008 \\
8473.82297 & 2.519 & 0.918 & 0.1690 & 0.008 \\
8473.82677 & 2.613 & 0.910 & 0.1705 & 0.008 \\
8474.83964 & 2.712 & 0.869 & 0.1770 & 0.008 \\
8474.84350 & 1.512 & 0.839 & 0.1654 & 0.008 \\
8475.84374 & 1.324 & 0.751 & 0.1586 & 0.008 \\
8475.84752 & 0.202 & 0.784 & 0.1609 & 0.008 \\
8476.82523 & -2.224 & 0.797 & 0.1631 & 0.008 \\
8476.82897 & 1.295 & 0.785 & 0.1571 & 0.008 \\
8479.84682 & -3.814 & 0.813 & 0.1623 & 0.008 \\
...&...&...&...&...\\
\hline
\end{tabular}
\end{table*}

\subsection{HARPS}
Eleven observations using HARPS were acquired between May 24 2018 and April 6 2019 from program IDs 0101.C-0497, 0102.C-0525 and 0103.C-0442 (PI: D\'iaz), in order to confirm the signal found in PFS data and also to constrain the orbital parameters of the planet candidate. The observations were carried out using simultaneous Thorium exposures with a fixed exposure time of 900 s reaching a mean signal-to-noise ratio of  $\sim$67 at 5500 \AA.
We re-processed the observations with the TERRA software \citep{AngladaEscudeButler2012}, where a high S/N template is constructed by combining all the observations that pass a threshold S/N cutoff, and then the RVs are computed by a $\chi^{2}$-fitting process relative to this template. The mean radial velocity uncertainty we get from this analysis is $\sim$0.89 \mps. TERRA also provides a computation of the S-indices and their uncertainties. These along with the RVs are listed in Table \ref{tab:harpsrvs}.

\begin{table*}
\center
\caption{TERRA Radial Velocities of HD 95338}
\label{tab:harpsrvs}
\begin{tabular}{lcccc}
\hline\hline
\multicolumn{1}{l}{BJD}& \multicolumn{1}{c}{RV}&\multicolumn{1}{c}{$\sigma$ RV}& \multicolumn{1}{c}{S}&\multicolumn{1}{c}{$\sigma$ S}\\ 
\multicolumn{1}{l}{(- 2450000)}& \multicolumn{1}{c}{(m s$^{-1}$)}&\multicolumn{1}{c}{(m s$^{-1}$)}& \multicolumn{1}{c}{(dex)}&\multicolumn{1}{c}{(dex)}\\ \hline 
8262.52210 & -2.347 & 0.963 & 0.1568 & 0.0016 \\
8263.58809 & -2.716 & 0.555 & 0.1642 & 0.0011 \\
8264.56962 & -2.820 & 0.775 & 0.1637 & 0.0014 \\
8265.60191 & -2.412 & 0.677 & 0.1672 & 0.0012 \\
8266.54165 & -4.199 & 1.105 & 0.1520 & 0.0018 \\
8429.84914 & 0.0 & 0.706 & 0.1580 & 0.0011 \\
8430.83705 & 1.651 & 0.712 & 0.1606 & 0.0009 \\
8576.69728 & 12.654 & 1.156 & 0.1584 & 0.0016 \\
8577.79238 & 14.113 & 1.479 & 0.1504 & 0.0023 \\
8578.71982 & 11.102 & 0.853 & 0.1564 & 0.0013 \\
8579.70958 & 11.115 & 0.790 & 0.1605 & 0.0012 \\
\hline
\end{tabular}
\end{table*}

\begin{figure}
 \includegraphics[width=\columnwidth]{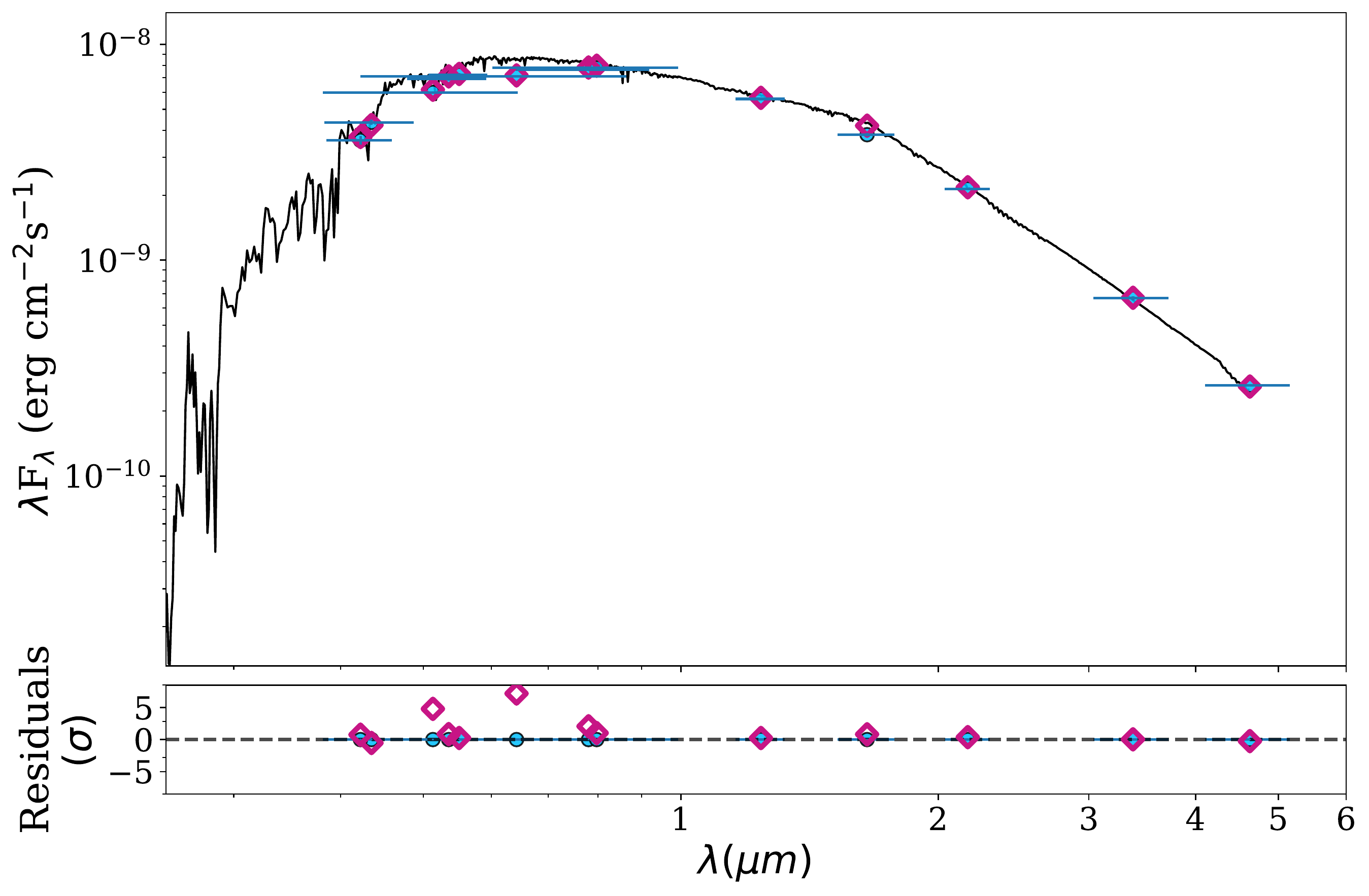}
 \caption{Top: best fitting BT-Cond SED model. Blue points are the photometry and magenta diamonds are the synthetic photometry. Horizontal error bars show the width of the filter bandpass. Bottom: Residuals of the fit, normalized to the photometry errors.}
 \label{fig:sed}
 \end{figure}
 \vspace{1cm}
  
 \section{Stellar Parameters}
We derived [Fe/H], $T_{\rm eff}$, age, mass, radius, log$g$ and $v$sin$i$ using the spectral classification and stellar parameter estimation package \texttt{SPECIES} \citep{Soto2018}, previously used in, e.g, \citet{Diaz2018,Diaz2020}. In short, \texttt{SPECIES} derives $T_{\rm eff}$, log\,$g$, [Fe/H] and microturbulence by measuring the equivalent widths (EWs) of a list of neutral and ionized iron lines, { and then using \texttt{MOOG} \citep{moog} to solve the radiative transfer equation in the stellar interior, along with ATLAS9 model atmospheres \citep{ATLAS9}. The adopted values for the atmospheric parameters are those for which no correlation is found between the individual iron abundance and the line excitation potential, nor the reduced EWs (EW/$\lambda$), and the average abundance for the FeI and FeII lines is the same.
The EWs used in this work were measured by fitting Gaussian-shaped profiles to the absorption lines through the \texttt{EWComputation}\footnote{Available at \url{https://github.com/msotov/EWComputation}} module in \texttt{SPECIES}. Details of the fitting procedure will appear in Soto et al. in prep. We produced a high signal-to-noise, stacked spectrum from HARPS observations to be used for the precise computation of the EWs. 
Physical parameters like mass and age are found by interpolation through a grid of MIST models \citep{Dotter2016}, using the \texttt{isochrones} python package \citep{Morton2015}. 
Finally, macroturbulence and rotation velocity were computed using temperature relations and fitting synthetic profiles to a set of five absorption lines (see \citealp{Soto2018} for more details).}

Then we performed a Spectral Energy Distribution (SED) fit to publicly available catalog photometry shown in Table \ref{tab:params} using the values found by \texttt{SPECIES} as priors. 

\begin{table}\label{tab:properties}
\center
\caption{Stellar Parameters of HD 95338.}
\label{tab:params}
\begin{tabular}{lcc}
\hline\hline
\multicolumn{1}{l}{Parameter}& \multicolumn{1}{c}{ Value  } & \multicolumn{1}{c}{Source}\\ \hline
\tess~ Name &TIC 304142124  &\\
R.A. (J2000)& 10:59:26.303  & SIMBAD\\
Dec. (J2000)& -56:37:22.947 & SIMBAD\\
\tess & 7.8436$\pm$0.0006& ExoFOP$^{a}$\\
$H $& 6.729$\pm$0.037 & 2MASS\\
$J$  & 7.098$\pm$0.024 & 2MASS\\
$Ks$& 6.591$\pm$0.017 & 2MASS\\
$V$ & 8.604$\pm$0.012 & Simbad\\
$B$ & 9.487$\pm$0.013&Simbad\\ 
$G$  & 8.3821$\pm$0.0003 & {\it Gaia}\\
$RP$& 7.8017$\pm$0.0013 &{\it Gaia}\\
$BP$& 8.8464$\pm$0.001 &{\it Gaia}\\
$W1$& 6.553$\pm$0.071 & Wise\\
$W2$& 6.578$\pm$0.023& Wise\\
Parallax (mas) &27.0553$\pm$0.0499& Gaia, \citet{Zinn2019}\\
Distance (pc) & 36.97$^{+0.02}_{-0.03}$ & This work\\ \hline
Spectral type & K0.5V & This work (\texttt{ARIADNE}) \\
Mass ($M_{\odot}$) & 0.83$^{+0.02}_{-0.02}$ & This work (\texttt{ARIADNE})\\
Radius ($R_{\odot}$) & 0.87$^{0.04}_{0.04}$ & This work (\texttt{ARIADNE})\\ 
Age (Gyr) & 5.08 $\pm$2.51 & This work (\texttt{SPECIES})\\
$A_{V}$ &0.073$^{+0.012}_{-0.015}$&This Work (\texttt{ARIADNE}) \\
Luminosity ($L_{\odot}$)& 0.49$\pm$0.01 & \citealt{Anderson2012}\\
T$_{\rm eff}$ (K) & 5212$^{+16}_{-11}$ & This work (\texttt{SPECIES})\\
\lbrack Fe/H\rbrack  &  0.04$\pm$0.10 & This work (\texttt{SPECIES})\\
log $g$& $4.54 \pm$ 0.21 & This work (\texttt{SPECIES})\\
$v$ sin $i$ (km s$^{-1}$) & 1.23 $\pm$ 0.28 & This work (\texttt{SPECIES})\\ 
$v_{\rm mac}$ (km s$^{-1}$) &0.97$\pm$0.41&This work (\texttt{SPECIES})\\ \hline

\multicolumn{3}{l}{$^{a}$\url{https://exofop.ipac.caltech.edu/tess/}}
\end{tabular}
\vspace{0.5cm}
\end{table}

The SED fit was done with \texttt{ARIADNE}, a python tool designed to automatically fit archival photometry to atmospheric model grids. \texttt{Phoenix v2} \citep{Husser2013}, \texttt{BT-Settl}, \texttt{BT-Cond} \citep{Allard2012}, \texttt{BT-NextGen} \citep{Hauschildt1999}, \cite{ATLAS9} and \cite{Kurucz1993} stellar atmosphere models were convolved with different filter response functions, $UBVRI$; 2MASS $JHK_\text{s}$ \citep{Skrutskie2006}; SDSS \textit{ugriz}; WISE $W1$ and $W2$; Gaia $G$, $RP$ and $BP$ \citep{Gaia2016,Gaia2018b}; Pan-STARRS $girwyz$; Str\"omgren uvby; GALEX NUV and FUV; {\it TESS}; {\it Kepler}; and NGTS to create 6 different model grids. We then model each SED by interpolating the model grids in $T_{\rm eff}-\log~{\rm g}-$[Fe/H] space. The remaining parameters are distance, radius, extinction in the $V$ band, and individual excess noise terms for each photometry point in order to account for possible underestimated uncertainties or variability effects. We set priors for $T_{\rm eff}$, $\log~{\rm g}$, and [Fe/H] from the \texttt{SPECIES} results, for the radius we took Gaia DR2 radius values as prior, for the distance we used the {\it Gaia} parallax as priors (after {applying the -52.8$\pm$2.4 $\mu$as correction from \citealp{Zinn2019}) and then we treated it as a free parameter in the fitting routine}. We limited the $A_{V}$ to a maximum of 4.243 taken from the re-calibrated SFD galaxy dust map (\citealt{Schlegel1998, Schlafly2011}). Each excess noise parameter has a zero mean Normal distribution as the prior, with the variance equal to five times the size of the reported uncertainty. We then performed the fit using \texttt{dynesty}'s nested sampler \citep{Speagle2019} to sample the posterior parameter space, obtaining the Bayesian evidence of each model and the marginalized posterior distribution for each fitted parameter as a by-product. Finally we averaged the posterior samples of each model, weighting each sample by its normalized evidence. To plot the SED, we selected the model grid with the highest evidence to calculate the synthetic photometry and overall model (Figure \ref{fig:sed}).  We note the residuals from Figure \ref{fig:sed} are normalized to the error of the photometry. In the case of precise photometry, e.g. {\it Gaia}, the residuals show a relatively high scatter. A more detailed explanation of the fitting procedure, accuracy, and precision of \texttt{ARIADNE} can be found in \citet{VinesJenkins2020}.

 \section{Detection from Radial Velocities}\label{sec:bayes}
\begin{figure}
 \includegraphics[width=\columnwidth]{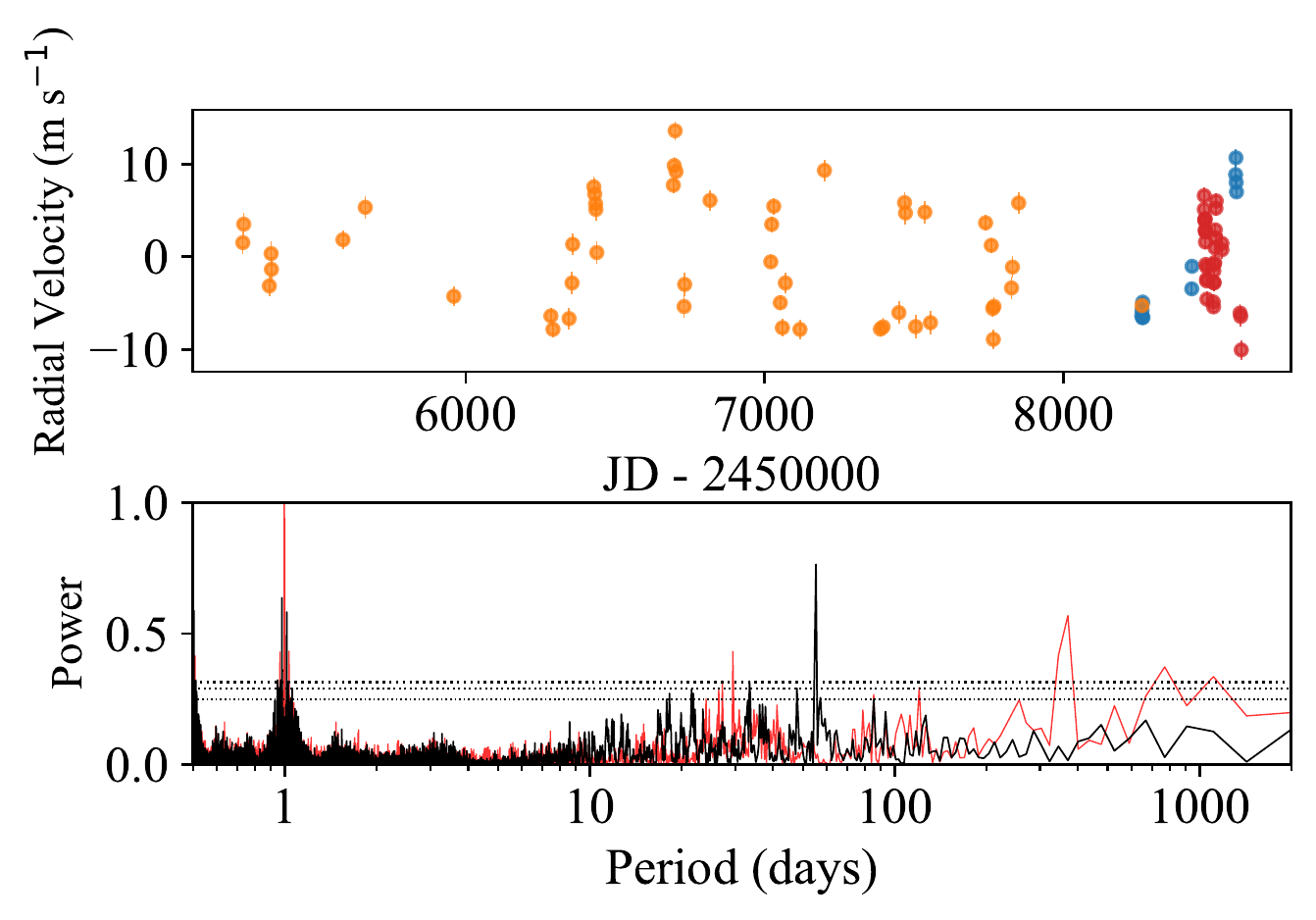}
 \caption{Top: Radial velocity time series for HD 95338 obtained with PFS1 (orange), PFS2 (red) and HARPS (blue). Bottom: GLS periodogram for the combined radial velocities. Each data set has been corrected by their respective velocity zero-point, estimated from the mean of the time series. Horizontal lines, from bottom to top, represent the 10, 1 and 0.1\% significance thresholds levels estimated from 5000 bootstraps with replacement on the data. The periodogram in red shows the window function for the time series.}
 \label{fig:rvper}
 \end{figure}

We began examining the radial-velocity data by using the traditional periodogram a\-na\-lysis approach to look for any periodicities embedded in the data. We used the generalized version \citep{Zechmeister2009} of the Lomb-Scargle periodogram (\citealt{Lomb1976, Scargle1982}, hereafter  GLS). Figure \ref{fig:rvper} shows the initial RV-only analysis where the signal at 55-days is clearly identified from the combined radial velocities. From this analysis we informed the following modeling process.

We modeled the radial velocities of HD 95338 following the same procedure defined in \citet{Tuomi2014} and performed in \citet{JenkinsTuomi2014} and \citet{Diaz2018} with some slight variations in our model. We define the global model as follows:

\begin{equation}
y_{i,j} = \hat{y}_{i,j} + \epsilon_{i,j} +\eta_{i,j} ~,
\label{eq:model}
\end{equation}
where
\begin{equation}\label{eq:rv}
    \hat{y}_{i,j}=\gamma_{j} + f_{k}(t_{i})
\end{equation}
is the deterministic part of the model composed of an offset $\gamma_{j}$ for data set $j$
and the Kepelerian component
\begin{equation}
 f_{k} (t_{i}) = \sum_{m=1}^{N_{p}} K_{m} [\, \text{cos}( \omega_{m}+ \nu_{m}(t_{i})) + e_{m} \text{cos}(\omega_{m}) ]~,
 \end{equation}
which is a function that describes a $m$-Keplerian model with $K_{m}$ being the velocity semi-amplitude, $\omega_{m}$ {\rm argument of periapsis of the star's orbit with respect to the barycenter}, $\nu_{m}$ is the true anomaly at the time of the planetary transit and $e_{m}$ is the eccentricity { for the $m$-th planet}. $\nu_{m}$ is also a function of the orbital period and the mean anomaly $M_{0,m}$, { measured at time $T_{0}$=2455253.72066.}
  
The stochastic component in the radial velocity data is modeled using a moving average (MA) approach, 
 \begin{equation}
\eta_{i,j}=\sum_{l=1}^{q}\phi_{j,l}\text{ exp }{\left\{ \frac{|t_{i-l} - t_{i}|}{\tau_{j}} \right\}} (v_{i-l,j}-\hat{y}_{i-l,j}) ~,
 \end{equation}
{ where $\phi_{j,l}$ represents the amplitude of the $q$th-order MA model, $\tau_{j}$ is the time scale of the MA($q$) model for the $j$-th instrument. The range of $\tau_{j}$ is determined according to the data timespan and cadence. Thus $\tau_{\rm max}$ = $t_{\rm max}$- $t_{\rm min}$, where $t_{\rm max}$ and $t_{\rm min}$ are the maximum and minimum value of the timespan of the combined set, respectively.   Finally, $\tau_{\rm min}$=min$\{t_2-t_1, t_3-t_2, ..., t_N-t_{N-1}\}$, represents the minimum difference between two epochs and $N$ is the total number of epochs.}
The white noise term in Equation \ref{eq:model} is denoted by $\epsilon_{i,j}$, where we assume that there is an excess white noise (jitter) in each data set with a variance of $\sigma_{j}$ such that $\epsilon_{i,j} \sim \mathcal{N}(0,\sigma^{2}_{i}+\sigma^{2}_{j})$, where $\sigma_{i}$ and $\sigma_{j}$ are the uncertainties associated with the measurement $y_{i,j}$ and jitter for the $j$-th dataset, respectively.

\subsection{Posterior Samplings and Signal Detection}\label{sec:tperi}
In order to estimate the posterior probability of the parameters in the model given the observed data we use Bayes' rule:
\begin{equation}\label{eq:bayesrule}
P(\theta\, | \,y) = \displaystyle \frac{ P(y\,| \,\theta)\, P(\theta) }{\,\int P(y \,| \,\theta)\,P(\theta)\,  d\theta}
\end{equation}

\noindent where $ P(y\,| \,\theta)$ is the likelihood function and $P(\theta)$ corresponds to the prior. The denominator is a normalizing constant such that the posterior must integrate to unity over the parameter space.  For our model, we choose the priors for the orbital and instrumental parameters as listed in Table \ref{tab:rvpriors}.

\begin{table}
\center
\caption{Prior selection for the parameters used in the MA analysis}
\label{tab:rvpriors}
\begin{tabular}{lcccc}
\hline\hline
\multicolumn{1}{l}{Parameter}&Units& \multicolumn{1}{c}{Prior Type}&\multicolumn{1}{c}{Range}\\ 
\hline
Semi-amplitude&m s$^{-1}$&Uniform& $K\in [\,0,100]\,$ \\
Logarithmic Period& day& Uniform &${\rm ln}P\in [\, {\rm ln}(1.1), {\rm ln}(10^{6})]\,$\\
Eccentricity& - & $\mathcal{N}(0,0.2)$ &$e \in [0,1)$\\
Long. of Peric.& rad &Uniform &$\omega \in [\, 0,2\pi]\,$\\
Mean Anomaly& rad &Uniform &$M_{0} \in [\, 0,2\pi]\,$\\
Jitter& m s$^{-1}$&Uniform & $\sigma_{J} \in [0,100] $\\
Smoothing time scale & day &Uniform & $\tau_{j} \in [\tau_{\rm min},\tau_{\rm max}]\,$ (see text)\\
MA Amplitude & - & Uniform & $\phi_{j} \in [\,0,1]\,$\\ 
\hline
\end{tabular}
\end{table}

For a given model, we sample the posterior through multiple tempered (hot) MCMC chains to identify the global maximum of the posterior. We then use non-tempered (cold) chains to sample the global maximum found by the hot chains. The procedure is similar to that previously done in \citet{Diaz2018} with the difference that here {our MA model includes a correlated (red) noise component but it does not include explicit correlations with activity indicators because it would introduce extra noise although it might remove some activity signals (see, e.g. \citealp{Feng2019b}). We explore the correlations between activity indices and radial velocities in Section \ref{sec:act_corr}. From the posterior samples, we infer the parameter at the mean value of the distribution and we report the uncertainties from the standard deviation of the distribution. This approach is also explained in detail in \citet{Feng2019a}.}
To select the optimal noise model, we calculate the maximum likelihood for a MA model using the Levenberg-Marquardt (LM) optimization algorithm \citep{Levenberg1944,Marquardt1963}. 

We define the Bayes Factor (BF) comparing two given models, $\mathcal{M}_{k}$ and $\mathcal{M}_{k-1}$, as 

\begin{equation}\label{eq:bayesfactor}
{\rm ln}\, B_{k,k-1} = {\rm ln}\, P(y | \mathcal{M}_{k}) - {\rm ln}\, P(y | \mathcal{M}_{k-1})
\end{equation}

We calculate ln(BF) for MA($q+1$) and MA($q$). If ln(BF)$< 5$, we select MA($q$), according to Equation \ref{eq:bayesfactor}. If ln(BF)$\geq$ 5, we select MA($q+1$) and keep increasing the order of the MA model until the model with the highest order passing the ln(BF)$\geq$ 5 criterion is found. Considering that the Bayesian information criterion (BIC) is a good criterion for signal selection \citep{KassRaftery1995, Feng2016}, we convert BIC to BF according to the formula given by \citet{Feng2016}. 

\begin{figure*}
\centering
\includegraphics[width=0.61\columnwidth]{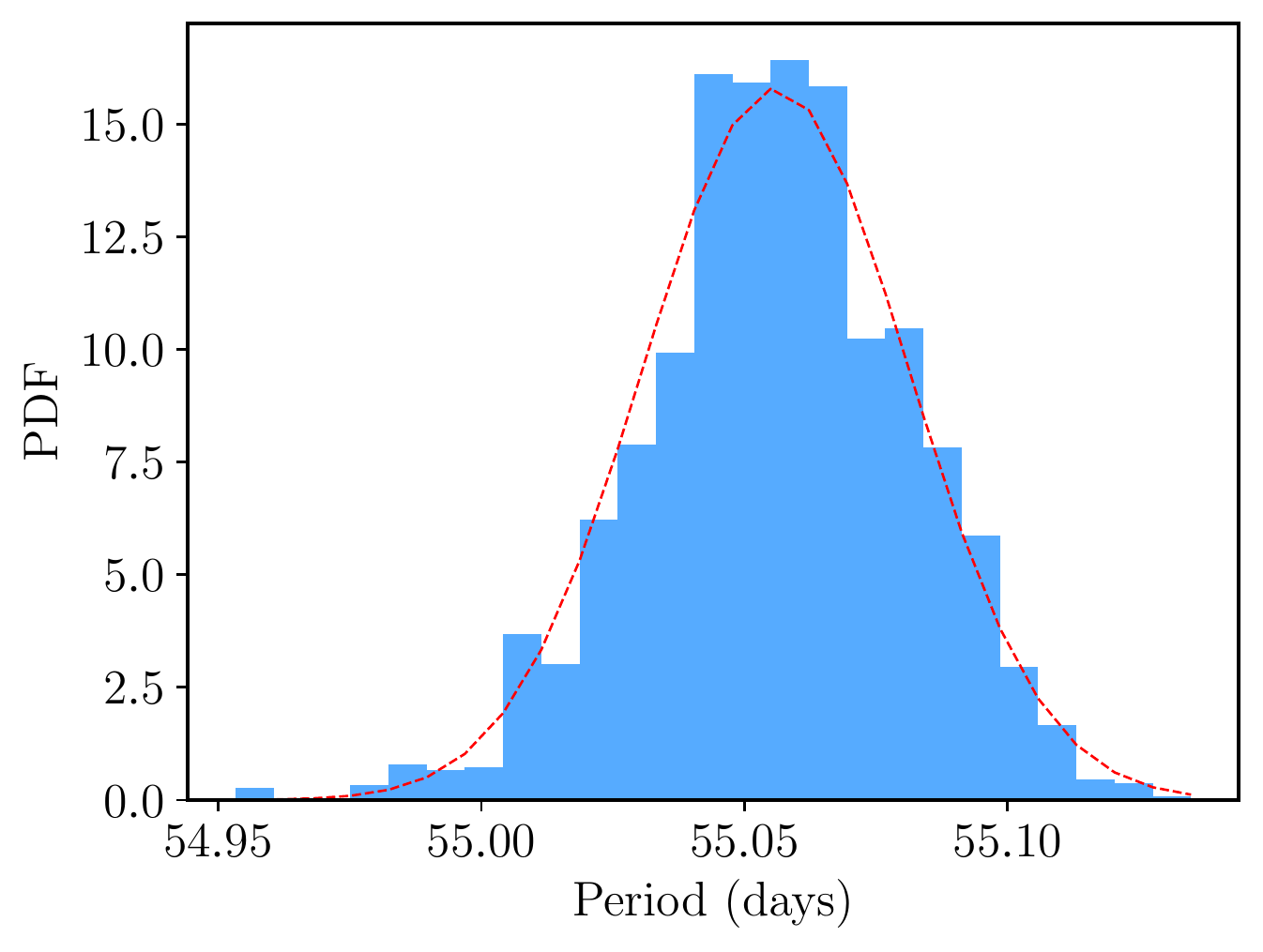}
\includegraphics[width=0.58\columnwidth]{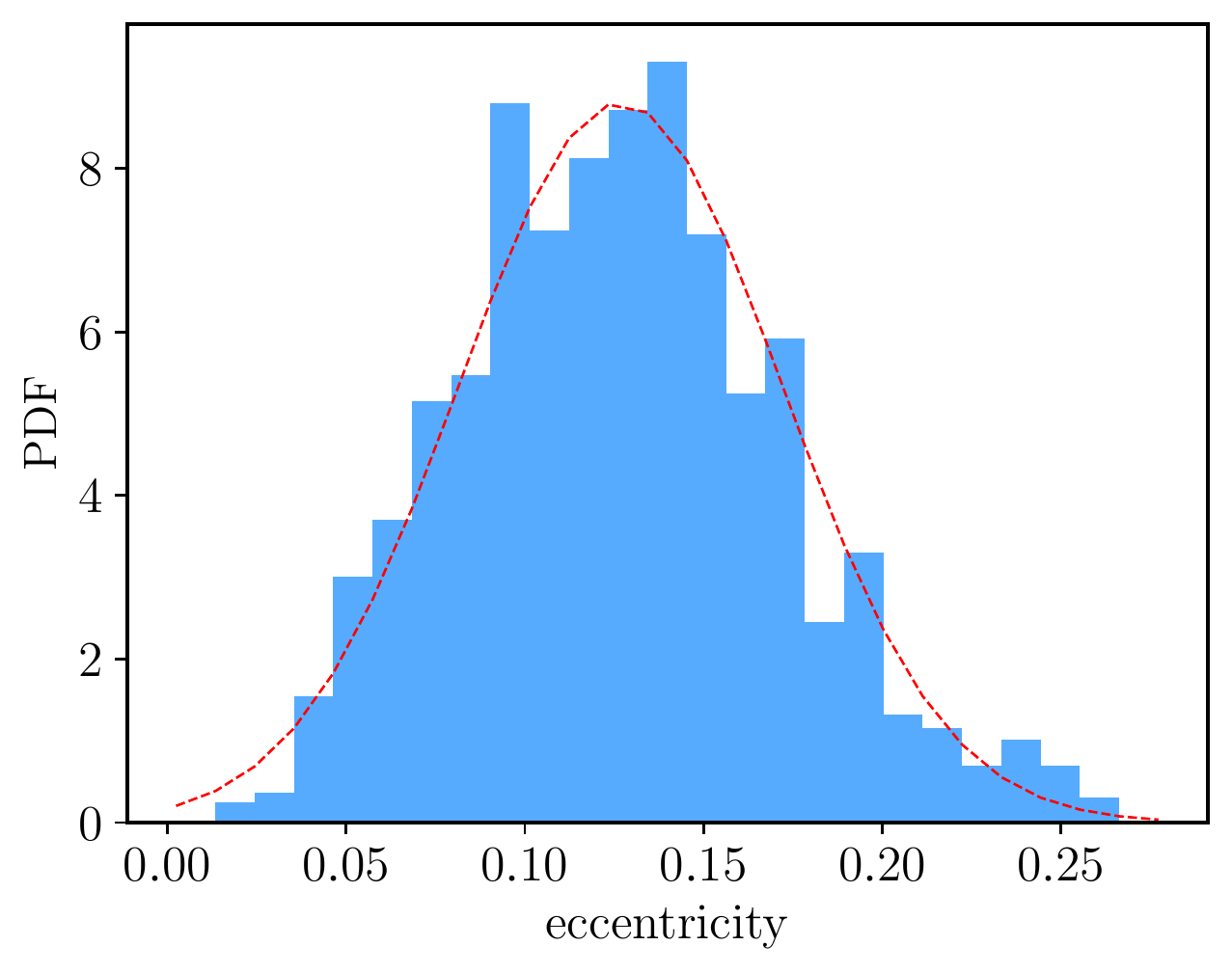}
\includegraphics[width=0.59\columnwidth]{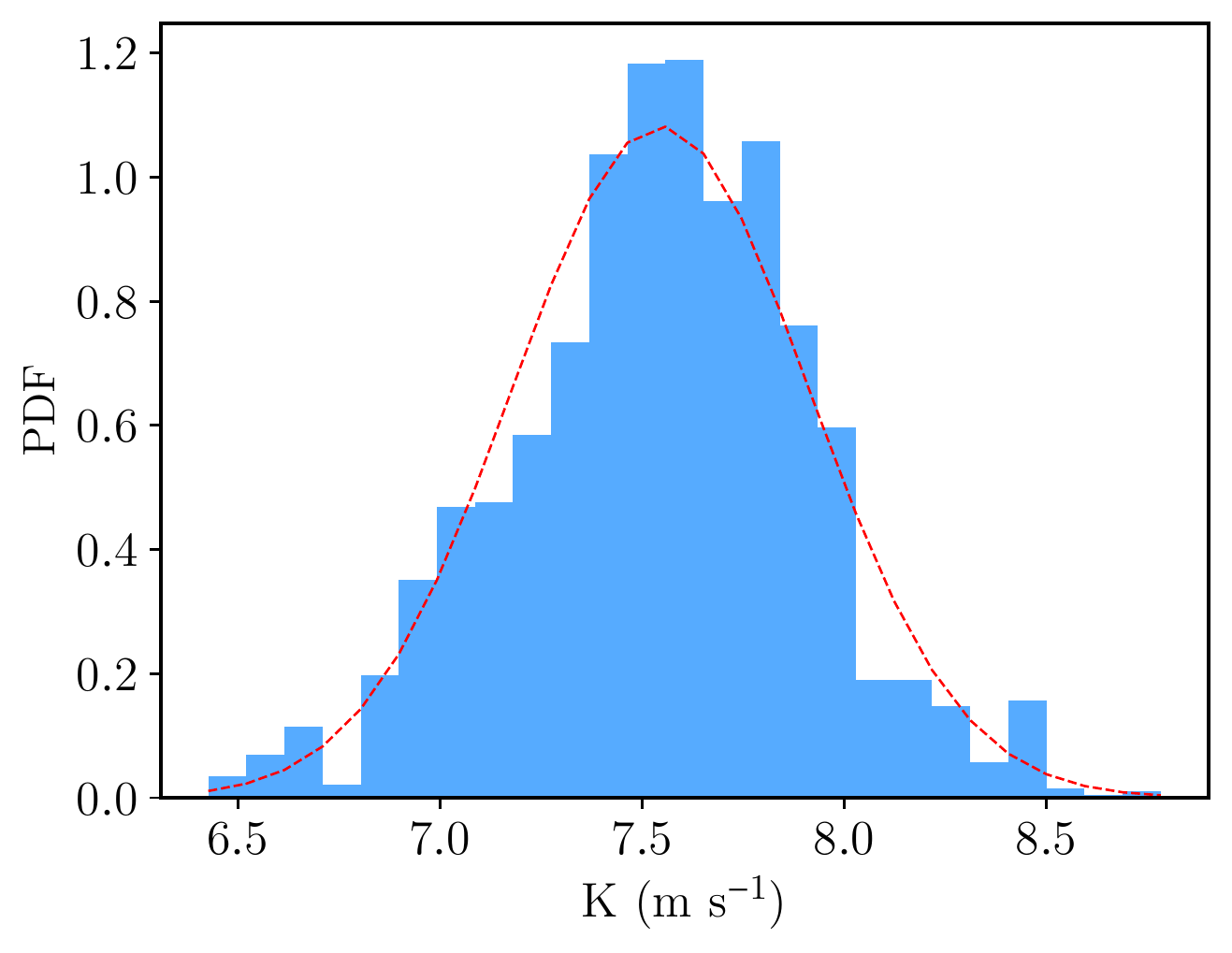}
\includegraphics[width=0.61\columnwidth]{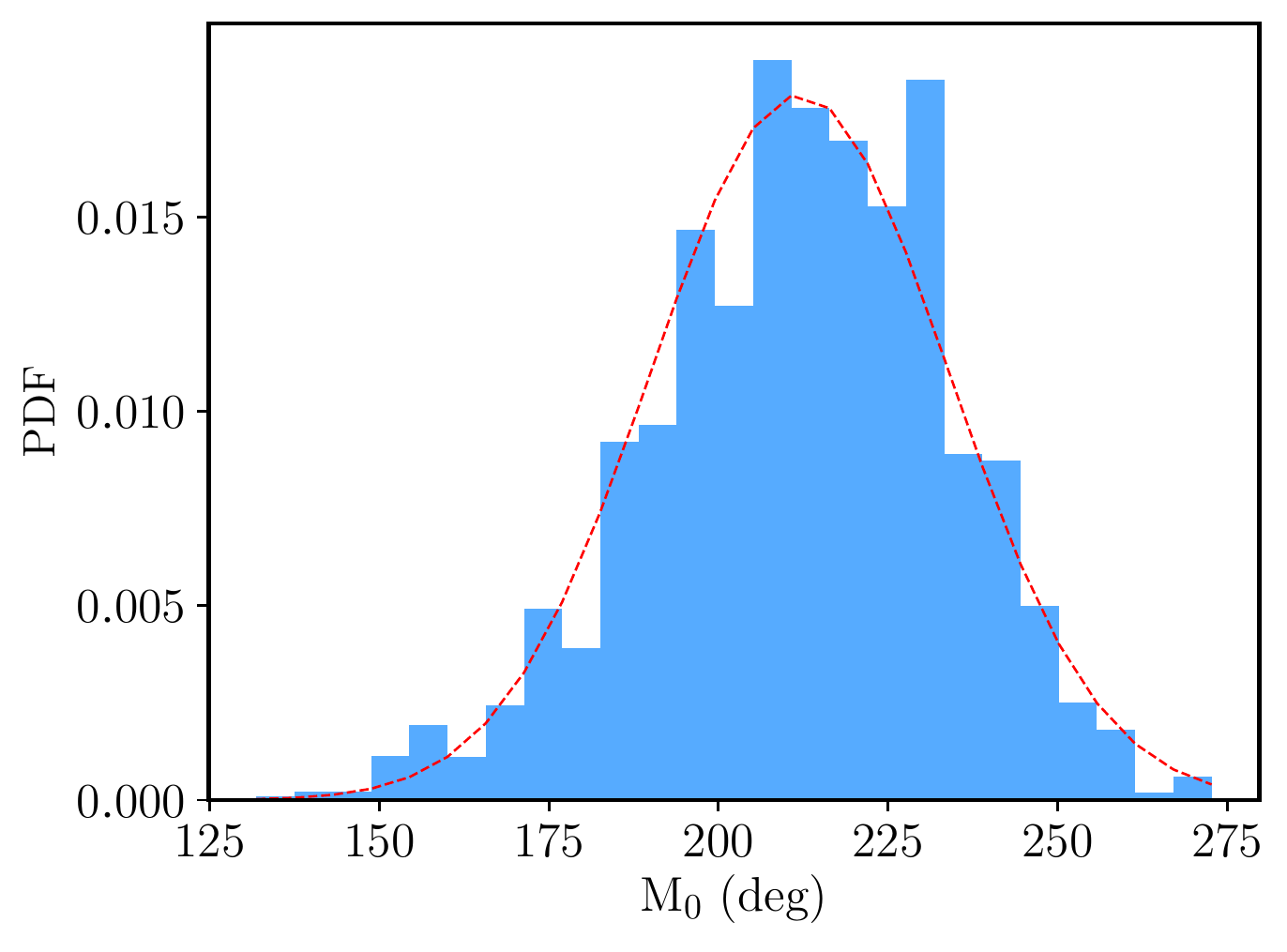}
\hspace{-0.1cm}\includegraphics[width=0.63\columnwidth]{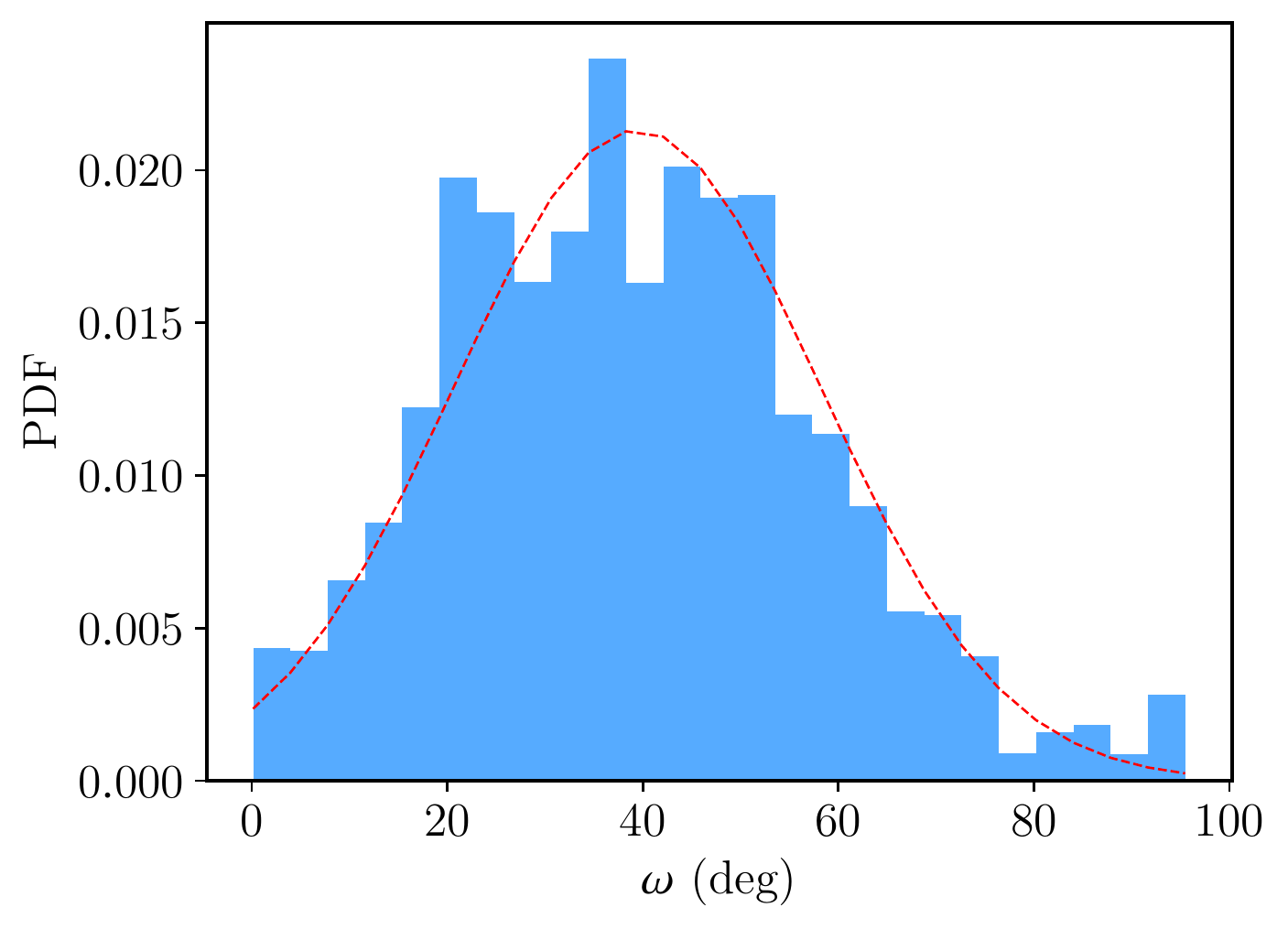}
\hspace{-0.15cm}\includegraphics[width=0.59\columnwidth]{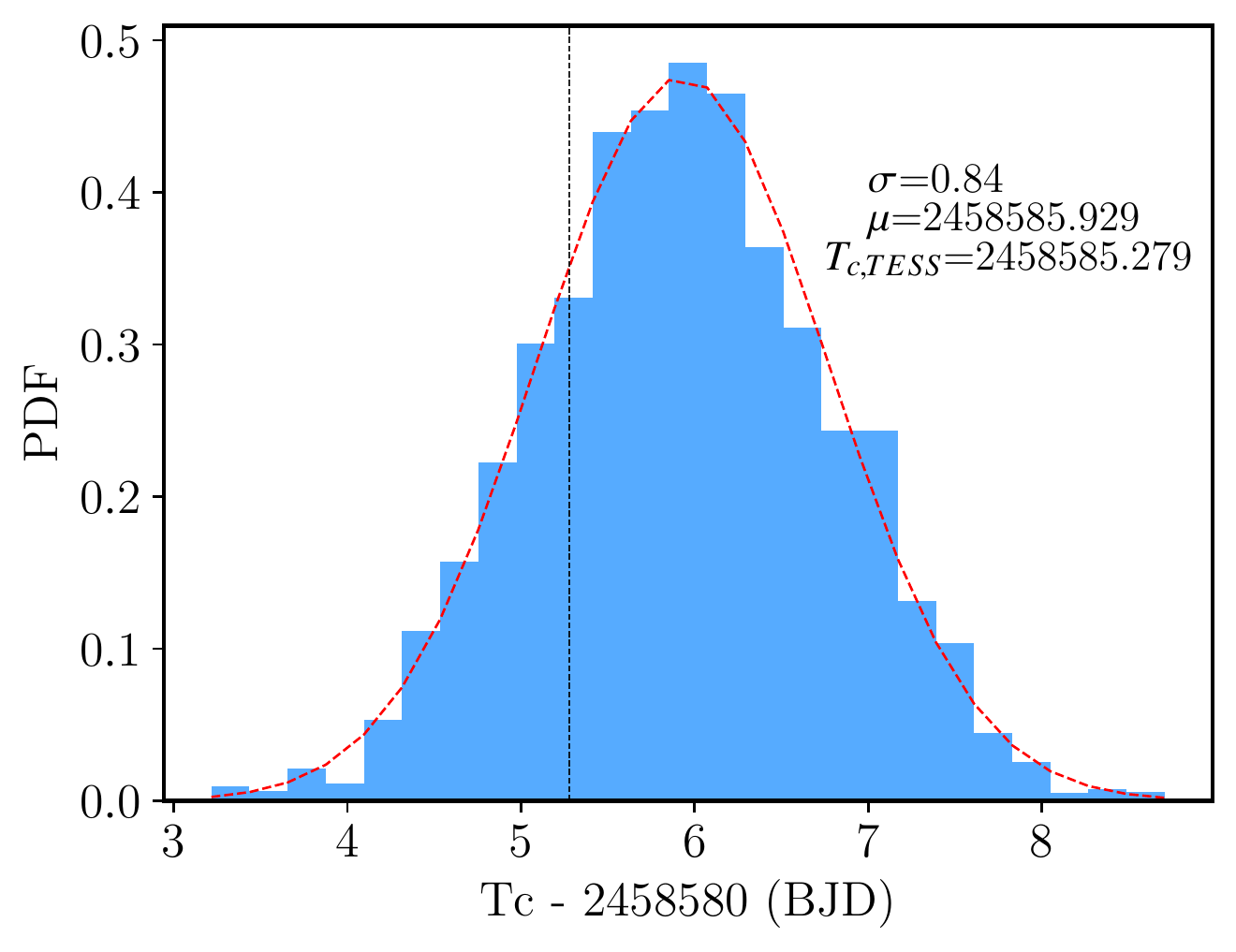}
\includegraphics[width=0.6\columnwidth]{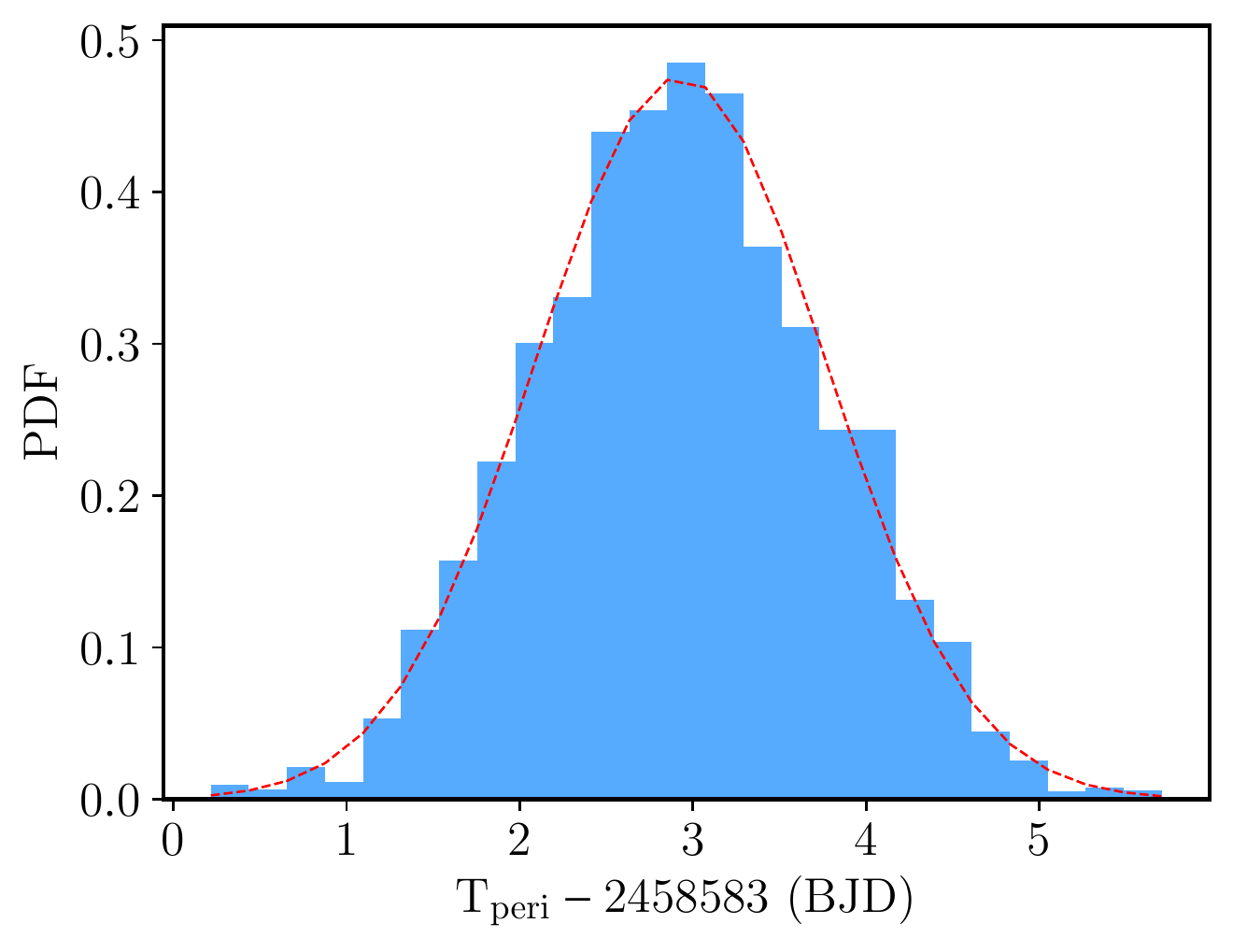}
\includegraphics[width=0.61\columnwidth]{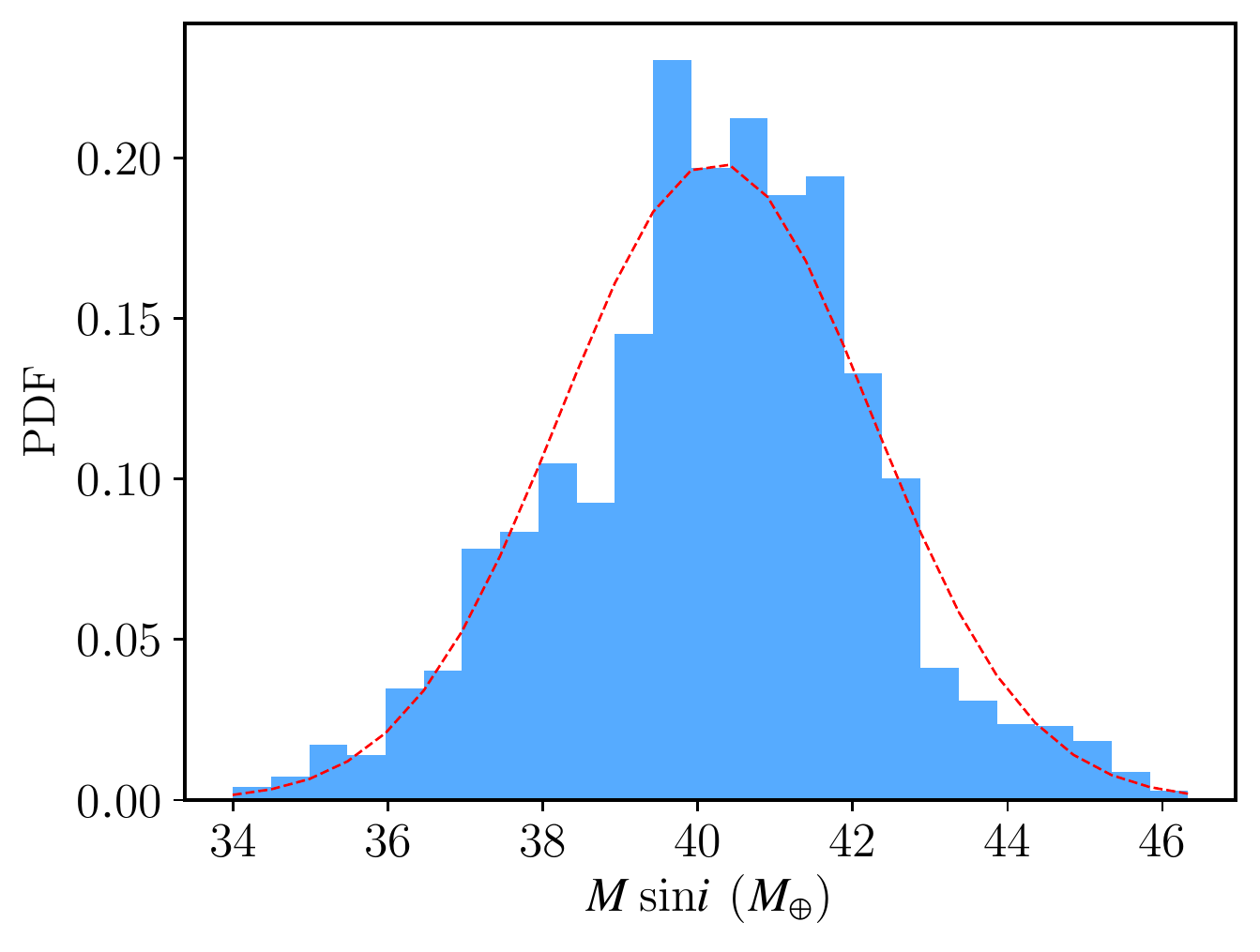}
\caption{Posterior distributions of the orbital parameters $P$, $e$, $K$, $M_{0}$, $\omega$, $T_{c}$, $T_{\rm peri}$ and minimum mass, respectively, obtained from our RV analysis. Dashed red lines on each plot show a Gaussian fit to the posterior distribution. $T_{c}$ is derived from the time of pericenter passage values ($T_{\rm peri}$, see text). Vertical black dashed line represents the transit time from the {\it TESS} lightcurve. From the histogram we found a mean value of $T_{c}=2458585.929$ and $\sigma$=0.84, which overlaps nicely with the transit time from the lightcurve, strongly suggesting both signals could originate from the same source.}
\label{fig:Tc}
\end{figure*}

{Our MCMC runs gave rise to the posterior histograms shown in Figure \ref{fig:Tc}, where the period, amplitude, and minimum mass (and the remaining orbital parameters) show Gaussian distributions centered on their respective mean values. }



From the posterior distributions for $T_{\rm peri}$ (see Figure \ref{fig:Tc}) we obtain $T_{c}=2458585.929\pm0.840$ which turns out to be well in agreement, within uncertainties, with the ephemeris from the \tess\, photometry, $T_{c,TESS}=2458585.279$ (see Table \ref{tab:planet}). {The posterior mean values for the radial velocity-only analysis are listed in Table \ref{tab:rvparams}.  It is worth noting that the final value for the timescale of the red noise, $\tau_{j}$, is not constrained for PFS2 as it did not converge to a unique solution.  We choose the best MA($q$) model based on 0-planet + MA($q$) model comparison and thus $q$ is determined based on the assumption that the time correlation in the RV data is totally noise, therefore $q$ is typically larger than it should actually be. This is the reason why the amplitude and time scale of MA($q$) models sometimes do not converge after adding Keplerian components which can explain the time correlation in the data better than stochastic red noise models such as MA. Although we can perform a selection of $q$ and number of signals simultaneously, it would be a 2-dimensional model selection and is thus time consuming. On the other hand, if a data set only contains white noise and signals, the Keplerian model will be favored against the MA model due to the advantage of simultaneous fitting. Compared to previous adoption of a single red noise model such as GP, our approach is more robust to overfitting or underfitting problems. }

\begin{table}
	\centering
	\caption{Posterior for the parameters included in the RV-only analysis.}
	\setlength{\tabcolsep}{10pt}
	\renewcommand{\arraystretch}{1.}
	\begin{tabular}{lc} 
\hline\hline
	Parameter 	&Value \\
	\hline
    $P$ (days)		&	55.056$\pm$0.025  \\
    $T_{\rm peri}$ (BJD - 2450000)&	8585.2795$\pm$0.8384\\
    $K$ (\ms) 	&7.54 $\pm$0.37	\\
    $e$ &0.127$\pm$0.045 \\
    $\omega$ (deg) &39.428 $\pm$ 18.719\\
    $M_{0}$ (deg) & 212.004$\pm{21.983}$\\
    $M {\rm sin }i$ (\me) &40.34$\pm$2.01  \\\hline
    $\mu_{\rm PFS1}$ (m s$^{-1}$)& 0.316$\pm$0.584 \\
    $\sigma_{J,\rm PFS1}$ (m s$^{-1}$)&1.725$\pm$0.818 \\
    $\phi_{\rm PFS1}$ &0.457 $\pm$0.426\\
    ln$\tau_{\rm PFS1}$ & 3.18$\pm$1.10\\
    $\mu_{\rm PFS2}$ (m s$^{-1}$)& 0.178$\pm$0.780 \\
    $\sigma_{J,\rm PFS2}$ (m s$^{-1}$)&0.985$\pm$0.532 \\
    $\phi_{\rm PFS2}$ &0.360 $\pm$0.314\\
    ln$\tau_{\rm PFS2}$ & 0.323$\pm$6.895\\
    $\mu_{\rm HARPS}$ (m s$^{-1}$)& 0.796$\pm$0.938 \\
    $\sigma_{J,\rm HARPS}$ (m s$^{-1}$)&1.80$\pm$0.87\\
    \hline
    \multicolumn{2}{l}{Note: MA(1) applied to PFS. White noise applied to HARPS.}
  	\end{tabular}
	    \label{tab:rvparams}

\end{table}
We note that additional tests were conducted using the Delayed Rejection Adaptive Metropolis algorithm \citep{Metropolis1953,Haario2001,Haario2006}, as previously done in \citet{Tuomi2014} and \citet{Diaz2018} and we found the results were in full agreement with the MA approach within the uncertainties.

\section{Stellar Activity and RV correlations}\label{sec:act_corr}
\begin{figure}
\centering
\includegraphics[width=\columnwidth]{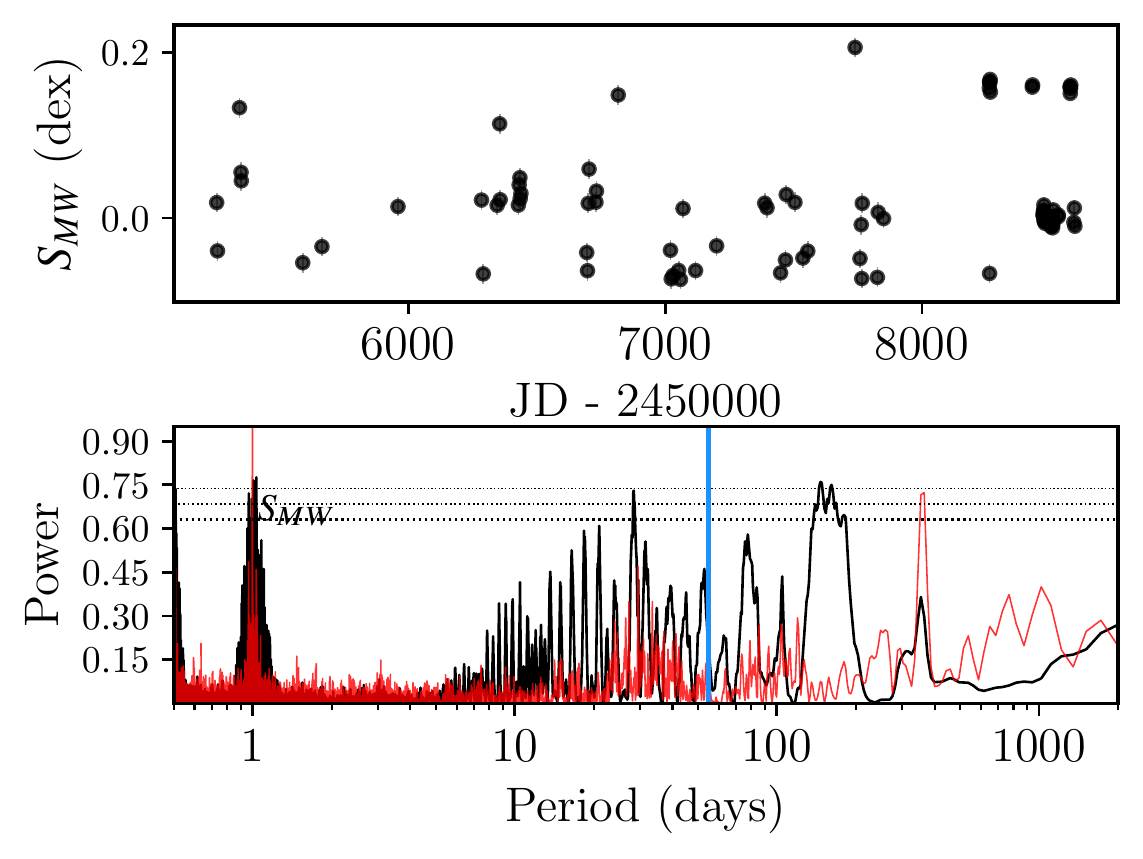}
\caption{Top: Time series of combined, mean subtracted S-indices from HARPS, PFS1 and PFS2. Bottom: GLS Periodogram of the S-indices. Vertical line shows the position of the 55-day radial velocity signal. Horizontal lines, from bottom to top, represent the 10, 1 and 0.1\% significance thresholds levels estimated from 5000 bootstraps with replacement on the data. }
\label{fig:actper}
\end{figure}

\begin{figure}
\label{fig:rvcor}
\centering
\includegraphics[width=\columnwidth]{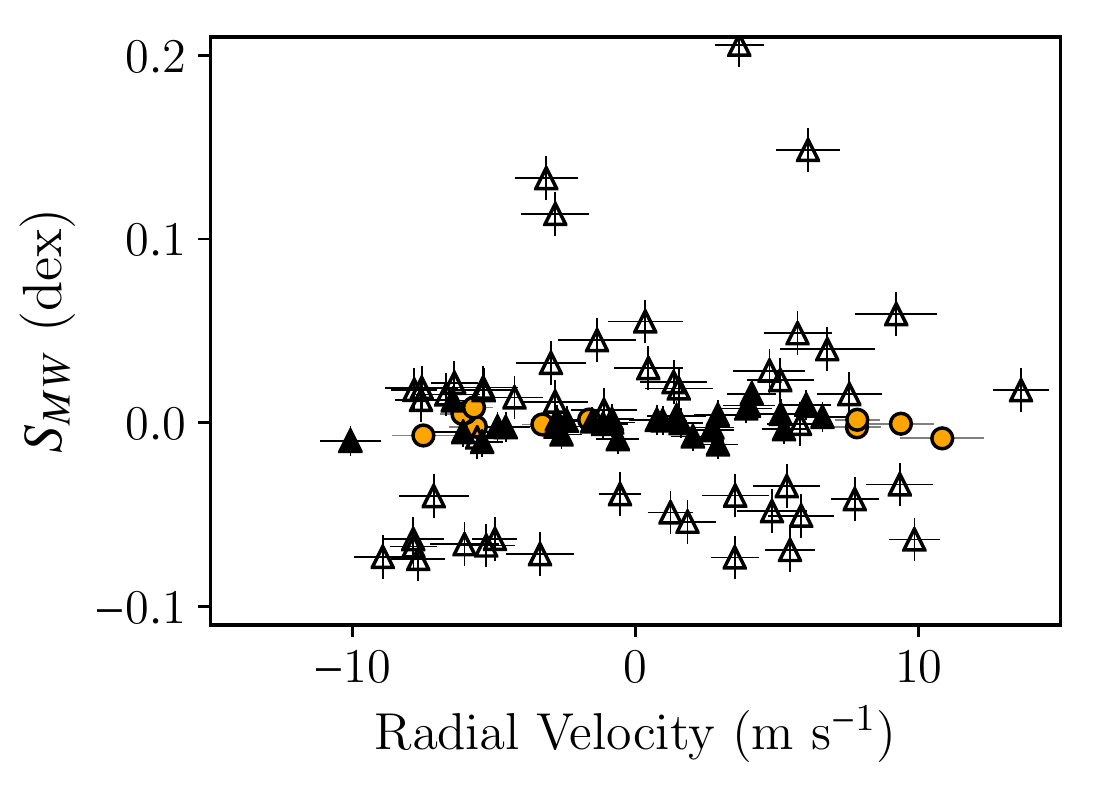}
\caption{Radial velocity correlations vs S-indices from HARPS (circles), PFS1(triangles), PFS2 (black triangles). }
\end{figure}

We computed the GLS periodogram of the combined S-indices from PFS1, PFS2 and HARPS (Figure \ref{fig:actper}). We do not find statistically significant periods from stellar activity matching the signal of the planet candidate (marked with a vertical line). However, we do see multiple peaks at $\sim$1, $\sim$29 and $\sim$150 days above the 1\% significance threshold. The 1-day period is likely due to the frequency of the sampling in the observations, similarly the 29~d peak is close to the lunar period. The additional 150~d period could be related to a stellar magnetic cycle, but more data is needed to test this hypothesis.
Figure \ref{fig:rvcor} shows the correlations between the mean-subtracted activity indices in the Mt. Wilson system, $S_{MW}$, and the radial velocities: PFS1 (open triangles), PFS2 (black triangles) and HARPS (orange circles). We note the improvement in the scatter from PFS2 compared to PFS1; new activity indices are comparable to the scatter of those from HARPS, derived using the TERRA software. We see 4 points that are far off from the mean. We find the Pearson $r$ correlation coefficients for PFS1, PFS2 and HARPS are 0.15, 0.38, -0.39, respectively, meaning no significant strong correlations are found ($\vert r \vert<0.5$)

\section{Photometry}
\subsection{TESS Photometry}

HD 95338 was observed by the Transiting Exoplanet Satellite Survey (\citealp[{\tess;}][]{Ricker2015}). We checked the target was observed using the Web {\it TESS} Viewing Tool (WTV\footnote{\url{https://heasarc.gsfc.nasa.gov/cgi-bin/tess/webtess/wtv.py}}), as initially the target did not produce an alert on the \tess~ Releases website\footnote{\url{https://tev.mit.edu/data/}} where an overview table, alerts and downloadable data is available. {We identified a single-transit in the {\tess} photometry containing data from Sector 10 using camera 3, observed between March 26th and April 22nd 2019.}

\begin{figure*}
\centering
\includegraphics[scale=0.75]{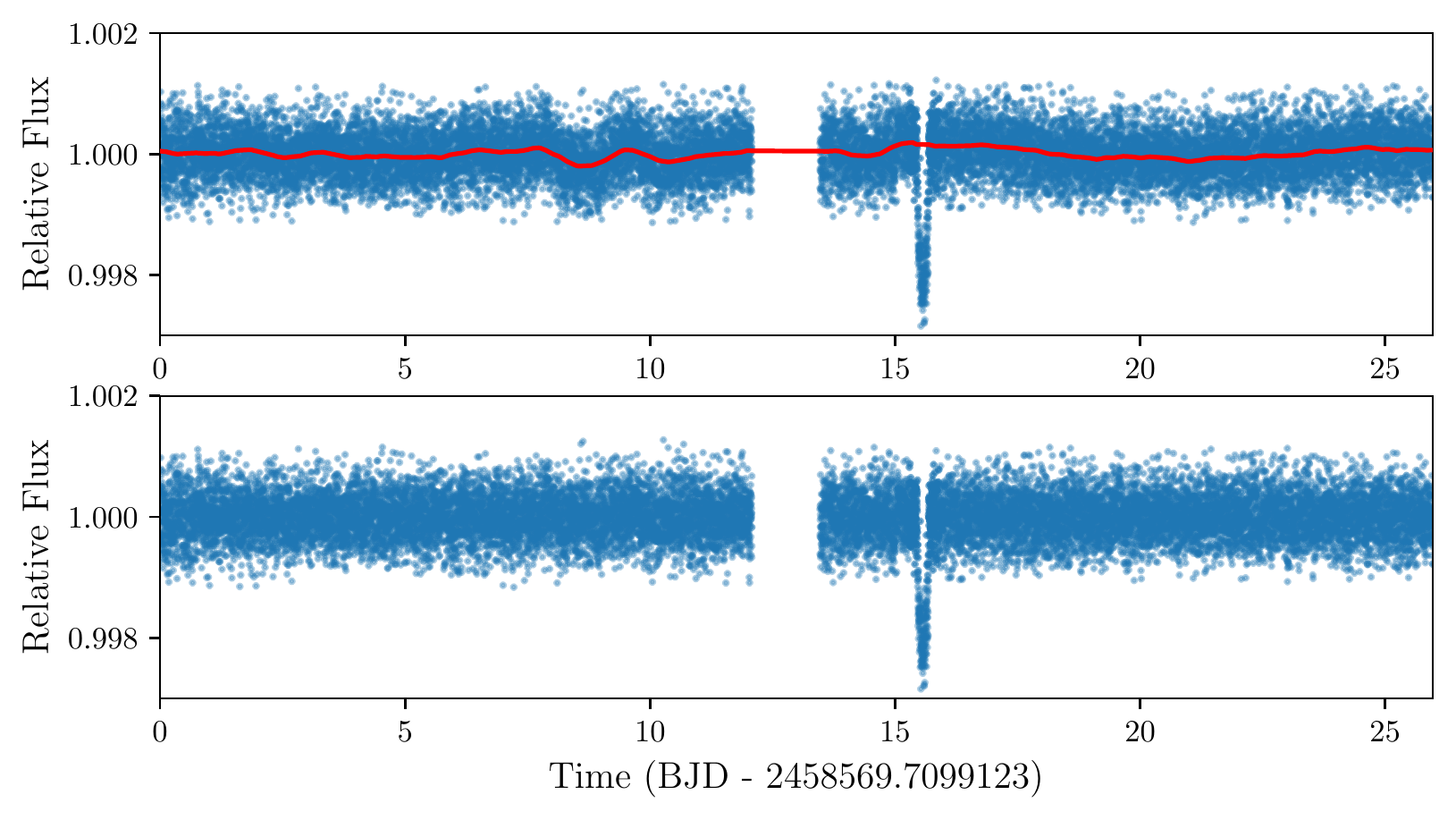}
\caption{Top: \texttt{PDC\_SAP} lightcurve for HD 95338 from {\it TESS} Sector 10 showing the single transit. Red solid curve on top of the photometry shows a median filter applied to remove variability. Bottom: Median filter corrected  \texttt{PDC\_SAP} TESS light curve for HD 95338.}
\label{fig:lc}
\end{figure*}

We extracted the \texttt{PDCSAP\_FLUX} 2-minute cadence photometry following the same procedures we recently used in \citet{Diaz2020}. The \texttt{PDCSAP\_FLUX}, median-corrected photometry is shown in the top panel of Figure \ref{fig:lc}. We then applied a median filter to remove the lightcurve variability, in particular on both sides near the transit event. The final flattened lightcurve is shown in the lower panel of Figure \ref{fig:lc} and it is the transit data used throughout all our analyses.

We note that the star is located in a relatively crowded field, as {\it Gaia} returns 12 sources within an angular separation of 1 arcmin. Given that the pixels in the \tess\, cameras are 21 arcsec wide, this could mean some of the sources would contaminate the aperture. However, the brightest nearby source is $G\sim$18 mag, which is 12 magnitudes fainter than HD 95338 ($G=8.38$). Converted into flux, this companion is $\sim$7,000 times fainter than HD 95338. {From a preliminary inspection and analysis of the light curve, we estimated a transit depth of $\sim$2000$\pm$500\footnote{https://exofop.ipac.caltech.edu/tess/target.php?id=304142124} ppm. Therefore, the difference in flux would cause a depth of $\sim100$ ppm, which we find to be negligible compared to the transit depth.}

Recent work by \citet{Sandford2019} have shown the use of single-transit lightcurves to estimate orbital periods based on precise parallaxes from Gaia. While their work focused on {\it K2} data, we can apply the same methodology to our {\it TESS} lightcurve, since we also know the transit depth, and we can calculate the scaled semi-major axis and stellar density from the combination of the \texttt{ARIADNE} results and the high resolution spectra. We recall equations 1 and 2 from \citet{Sandford2019}:

\begin{equation}
P^{2} = \frac{3\pi}{G}\Big(\frac{a}{R_{\star}}\Big)^{3}\rho_{\star}^{-1}
\end{equation}\label{eq:period}

\begin{equation}
\sigma_{P} = \frac{P}{2}\sqrt{\Big( \frac{\sigma_{\rho_{\star}}}{\rho_{\star}}\Big)^{2}+ \Big( \frac{3\sigma{\frac{a}{R_{\star}}}}{\frac{a}{R_{\star}}}\Big)^{2}}
\end{equation}\label{eq:err_period}
which yield the orbital period (and the associated error) of a single transit using Kepler's third law and assuming circular orbits, where $G$ is the gravitation constant, $(a/R_{\star})$ corresponds to the scaled semi-major axis measured directly from the shape of the transit and $\rho_{\star}$ is the stellar density that must come from an independent analysis. In our case, we used the stacked spectra acquired with HARPS, and from our spectra classification analysis with \texttt{SPECIES} combined with the SED fit, we find a stellar density of $\rho_{\star}$=1.68$^{+0.45}_{-0.23}$ \gcm. We estimate $(a/R_{\star}$)= 58.06$^{+1.39}_{-2.48}$ from the transit seen in the {\it TESS} lightcurve. Then, using equations (1) and (2) from from \citet{Sandford2019} we get an estimate for an orbital period of 47$\pm$9 days for the single transit observed by {\it TESS} being consistent within the uncertainties to the period of the signal found in the radial velocity data.

\subsection{ASAS Photometry}\label{sec:photo}
\begin{figure}
\centering
\includegraphics[width=\columnwidth]{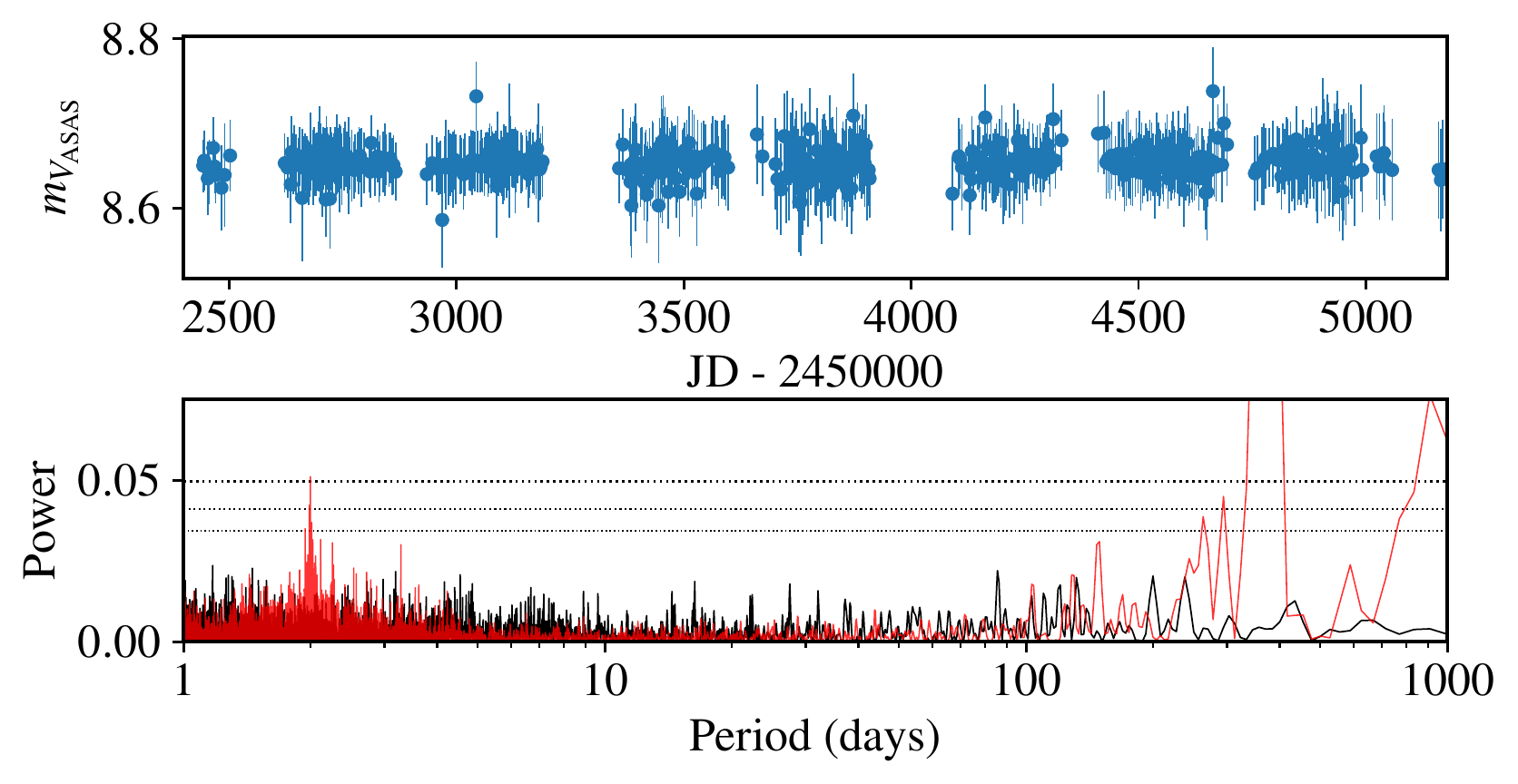}
\caption{GLS periodogram for the ASAS V-band photometry. Horizontal lines mark the position of the 10,1 and 0.1\% FAP threshold levels, from bottom to top, respectively. A peak close to $\sim$90 days is seen in the power spectrum, however it is below any FAP threshold and cannot be considered as statistically significant.}
\label{fig:asas}
\end{figure}

In an attempt to search for additional sources of periodicity we used data from the All Sky Automated Survey (\citealp[ASAS;][]{Pojmanski1997}). Figure \ref{fig:asas} shows the photometry time series consisting on 625 measurements from December 7th 2000 to December 3rd 2009. We selected the best quality data, flagged as ``A" or ``B". We used the GLS periodogram to search for signals after filtering the highest quality data from outliers, and found no statistically significant periods that could be attributed to the stellar rotation period, due in part to the size of the typical uncertainty in the ASAS photometry. \newline

\begin{table*}
    \centering
       
   \caption{Priors used on the joint analysis of HD 95338.}
     \label{tab:priors}
    \setlength{\tabcolsep}{5.5pt} 
    \begin{tabular}{lccl} 
        \hline
        \hline
        Parameter name & Prior & Units & Description \\
                \hline
                \hline        
         ~~~$\rho_{\star}$ & $\mathcal{N}$(1685,30)& kg m$^{-3}$ & Stellar density. \\
        Parameters for planet b\\
	~~~$P_b$ & $\mathcal{J}(1,100)$ & days & Orbital Period. \\
        ~~~$T_{c,b} - 2457000$ & $\mathcal{U}(1000,1100)$  & days & Time of transit-center. \\
        ~~~$r_{1,b}$ & $\mathcal{U}(0,1)$ & --- & Parametrization for $p$ and $b$${}^{1}$. \\
        ~~~$r_{2,b}$ & $\mathcal{U}(0,1)$ & --- & Parametrization for $p$ and $b$${}^{1}$. \\
        ~~~$K_{b}$ & $\mathcal{U}(1,100)$ & m s$^{-1}$ & Radial-velocity semi-amplitude. \\
       ~~~${e}_{b}$ & $\mathcal{U}(0,1) $ & ---& eccentricity.\\
       ~~~$\omega_{b}$& $\mathcal{U}(0,359.)$ & deg & argument of periastron.\\
        \hline
        Parameters for TESS\\
        ~~~$D_{\textnormal{TESS}}$ &  1.0 (Fixed)  & --- & Dilution factor for \textit{TESS}. \\
        ~~~$M_{\textnormal{TESS}}$ & $\mathcal{N}(0,1000)$ & ppm & Relative flux offset for \textit{TESS}. \\
        ~~~$\sigma_{w,\textnormal{TESS}}$ & $\mathcal{J}(0.1,100)$ & ppm & Extra jitter term for \textit{TESS} lightcurve. \\
        ~~~$q_{1,\textnormal{TESS}}$ & $\mathcal{U}(0,1)$ & --- & Quadratic limb-darkening parametrization. \\
        ~~~$q_{2,\textnormal{TESS}}$ & $\mathcal{U}(0,1)$ & --- & Quadratic limb-darkening parametrization. \\
        \hline
        RV instrumental parameters\\
        ~~~$\mu_{\textnormal{PFS1}}$ & $\mathcal{N}(0,10)$ &m s$^{-1}$ & Radial velocity zero-point (offset) for PFS1.\\
        ~~~$\sigma_{w,\textnormal{PFS1}}$ & $\mathcal{J}(0.1,10)$ & m s$^{-1}$ & Extra jitter term for PFS1 radial velocities. \\
        ~~~$\mu_{\textnormal{PFS2}}$ & $\mathcal{N}(0,10)$ &m s$^{-1}$ & Radial velocity zero-point (offset) for PFS2.\\
        ~~~$\sigma_{w,\textnormal{PFS2}}$ & $\mathcal{J}(0.1,10)$ & m s$^{-1}$ & Extra jitter term for PFS2 radial velocities. \\
        ~~~$\mu_{\textnormal{HARPS}}$ & $\mathcal{N}(0.,10)$ &m s$^{-1}$ & Radial velocity zero-point (offset) for HARPS.\\
        ~~~$\sigma_{w,\textnormal{HARPS}}$ & $\mathcal{J}(0.1,10)$ & m s$^{-1}$ & Extra jitter term for HARPS radial velocities. \\
        \hline
        \multicolumn{3}{l}{$^{1}$We used the transformations outlined in \citet{Espinoza2018efficient}.}\\
         \end{tabular}
\end{table*}

\begin{table}
	\centering
	\caption{Planetary Properties for HD 95338\,b}
	\setlength{\tabcolsep}{10pt}
	\renewcommand{\arraystretch}{1.2}
	\begin{tabular}{lc} 
\hline\hline
	Property	&	Value \\
	\hline\hline
    Fitted Parameters &\\
    ~~~$\rho_{\star}$ (kg m$^{-3})$ & 1686.537$^{+29.810}_{-29.993}$ \\ \hline
    ~~~$P$ (days)		&	55.087$^{+0.020}_{-0.020}$  \\\vspace{0.5pt}
    ~~~$T_{c}$ (BJD - 2450000)&	8585.2795$^{+0.0006}_{-0.0006}$	\\\vspace{0.5pt}
    ~~~$a/R_{*}$ &64.676$^{+0.381}_{-0.384}$ \\\vspace{0.5pt}
    ~~~$b$ & 0.430$^{+0.070}_{-0.113}$    \\\vspace{0.5pt}
    ~~~$K$ (m s$^{-1}$) 	&8.17 $^{+0.42}_{-0.39}$	\\
    ~~~$i_{\rm p}$ (deg)  & 89.57$^{+0.09}_{-0.05}$ \\\vspace{0.5pt} 
    ~~~$e$ &0.197$^{+0.029}_{-0.024}$ \\
    ~~~$\omega$ (deg) & 23.42$^{+11.53}_{-11.99}$\\\hline
    Derived Parameters & \\
    ~~~$M_{\rm p}$ (\me) &42.44$^{+2.22}_{-2.08}$  \\\vspace{0.5pt}
    ~~~$R_{\rm p}$ (\re)& 3.89$^{+0.19}_{-0.20}$ \\\vspace{0.5pt}
    ~~~$a$ (AU) & 0.262$^{+0.002}_{-0.002}$ \\\vspace{0.5pt}
    ~~~$\rho_{\rm p}$ (\gcm) & 3.98$^{+0.62}_{-0.64}$ \\\vspace{0.5pt}
    ~~~$T_{\rm eq}^{1}$ (K) &385$^{+17}_{-17}$ \\
    ~~~$\langle F \,\rangle$ ($\times$10$^{7}$ erg s$^{-1}$ cm$^{-2}$)& 1.01$\pm$0.03\\
    \hline
    Instrumental Parameters &\\
    ~~~$M_{\textnormal{TESS}}$ (ppm) & -0.0000027$^{+0.0000028}_{-0.0000027}$\\
    ~~~$\sigma_{w, \rm TESS}$ (ppm) & 1.836$^{+12.323}_{-1.570}$\\\vspace{0.5pt}
    ~~~$q_{1,\rm TESS}$ & 0.389$^{+0.109}_{-0.073}$ \\\vspace{0.5pt}
    ~~~$q_{2,\rm TESS}$ & 0.848$^{+0.108}_{-0.183}$\\\vspace{0.5pt}
    ~~~$\mu_{\rm PFS1}$ (m s$^{-1}$)& 0.77$^{+0.36}_{-0.35}$ \\\vspace{0.5pt}
    ~~~$\sigma_{w,\rm PFS1}$ (m s$^{-1}$)&2.31$^{+0.32}_{-0.28}$ \\
    ~~~$\mu_{\rm HARPS}$ (m s$^{-1}$)& 3.83$^{+0.59}_{-0.56}$ \\\vspace{0.5pt}
    ~~~$\sigma_{w,\rm HARPS}$ (m s$^{-1}$)&1.61$^{+0.54}_{-0.40}$ \\
    ~~~$\mu_{\rm PFS2}$ (m s$^{-1}$)& -1.01$^{+0.27}_{-0.28}$ \\\vspace{0.5pt}
    ~~~$\sigma_{w,\rm PFS2}$ (m s$^{-1}$)&1.30$^{+0.30}_{-0.26}$ \\
    \hline
    \multicolumn{2}{l}{$^{1}$Estimated using a Bond albedo of 0.5.}
	\end{tabular}
	    \label{tab:planet}
\end{table}

{In order to address how often we could recover a prediction for the transit centroid, $T_{c}$, that has an uncertainty of 1.5\% of the orbital period or better, just as we see for HD 95338 b, we simulated 10$^{6}$ systems with a single planet and random orbital parameters. We consider that all the random systems transit their host stars and we used flat priors for the distribution of longitude of pericenter, $\omega$, and for the eccentricity. For the distribution of orbital periods we used the broken power law presented in \citet{Mulders2018b}, where the break occurs at $P_{b}$=10 days. For shorter periods the probability is written as $(P/P_{b})^{1.5}$, while for longer periods the probability is unity. For each system, we generated the remaining orbital parameters according to standard equations for the orbital parameters, use these to predict $T_{c}$ (see Section \ref{sec:bayes}). We find that $\sim$9\% of the systems sampled randomly fulfill this criterion. 

If the agreement between the RV prediction and transit $T_{c}$ found for HD 95338 is just a statistical fluke, then this means there are more planets in the system, since another body must give rise to the transit.  The probability of 9\% does not consider this possibility.  For that to be the case, we should also normalize by the fraction of Neptunes that are found in multiple systems. Although this value is uncertain, and may actually be $\sim$100\%, we can at least estimate it using a literature search. To do this, we retrieved the number confirmed Neptunes with known companions detected by the transit method by {\it Kepler/K2} from the \texttt{exoplanet.eu}\footnote{\url{http://exoplanet.eu/catalog/}} catalog in a mass range between 10 and 45 $M_{\oplus}$. We find that the number of these multi-systems is 19 out of a total of 65, which corresponds to a fraction of $\sim$29\%. This leads to a final probability of $\sim$3\%, meaning it is highly unlikely that we have observed the configuration we find for HD 95338 b if the orbital parameters are randomly distributed.  Even if Neptunes are indeed found to exist exclusively in multi-planet systems, there is still a 91\% probability that the RV detected companion and the \tess\, detected companion are the same object. }

\section{Joint Analysis}
{We performed a joint fit of the photometry and radial velocities (Tables \ref{tab:pfs1rvs} to \ref{tab:harpsrvs}) using the \texttt{juliet}\, package \citep{Espinoza2019b} in order to estimate the orbital parameters for the system. To model the photometry \juliet~ uses the \texttt{batman} package \citep{Kreidberg2015} while the radial velocities are modeled using \texttt{radvel} \citep{Fulton2018}. We then sampled the parameter space using the \texttt{dynesty} nested sampler \citep{Speagle2019} to compute posterior samples and model evidences. The parameters for the joint model were set according to Table \ref{tab:priors}. }
We treated the eccentricity as a free parameter motivated by our finding from the RV-only analysis suggesting the eccentricity was different from zero. The resultant value was in agreement with the one from our previous analysis. {The RV semi-amplitude prior was chosen to be flat between 1 and 100 to explore a wider range of amplitudes and not only values centered around the semi-amplitude found in the RV-only analysis. The jitter terms for PFS1, PFS2 and HARPS, were set using a Jeffreys prior over two orders of magnitude (0.1 to 10 \mps)}, resulting in excess RV noise of 2.3, 1.3 and 1.6 \mps, respectively. 
{For the orbital period we used a Jeffreys prior over two orders of magnitude, from 1 to 100 days.} The time of transit ($T_c$) was derived from the time of pericenter pasage ($T_{\rm peri}$) as discussed in Section \ref{sec:tperi}. {However, we also chose an uninformative prior using the whole range of the radial velocity baseline. }

\begin{figure*}
\includegraphics[scale=0.67]{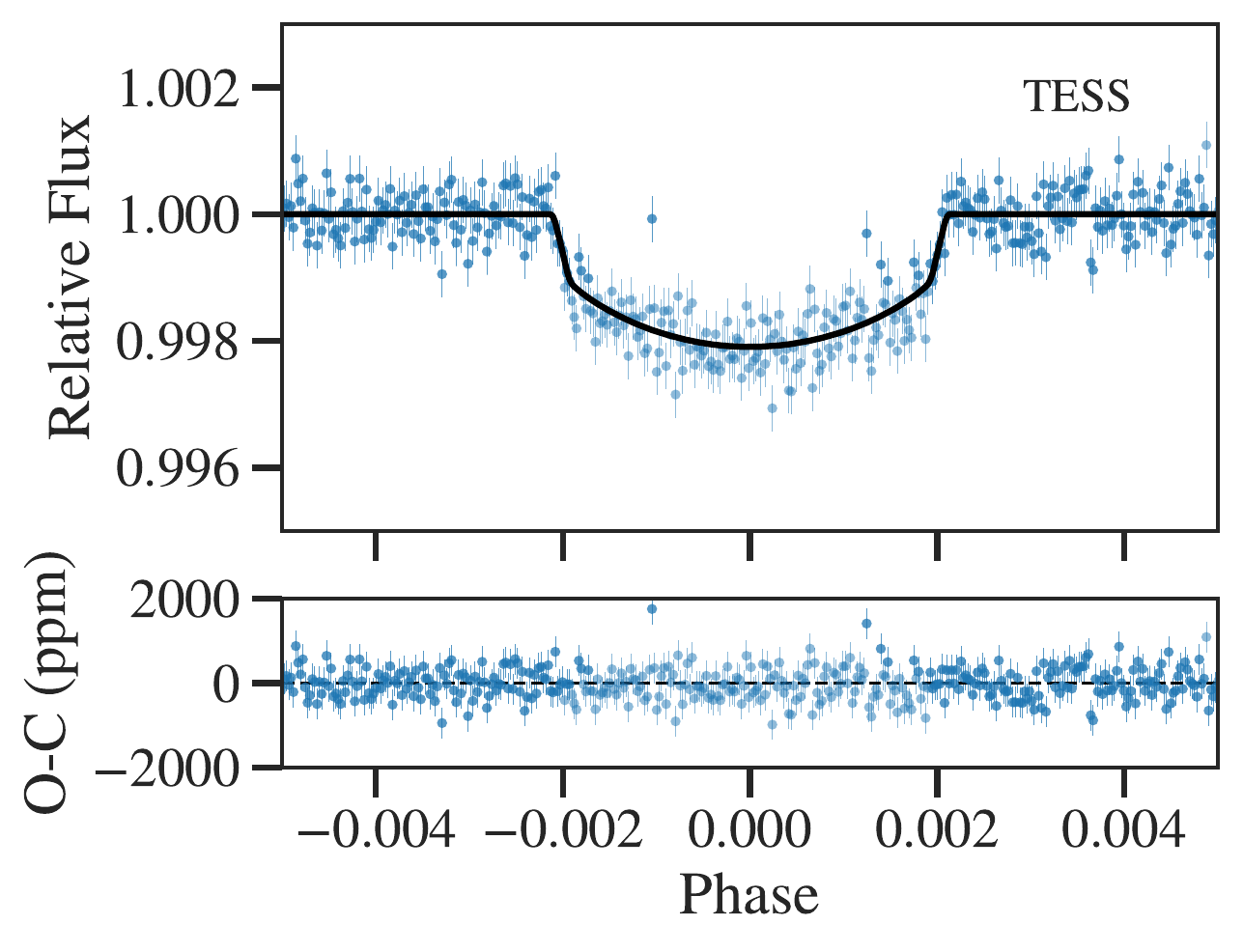}
\includegraphics[scale=0.656]{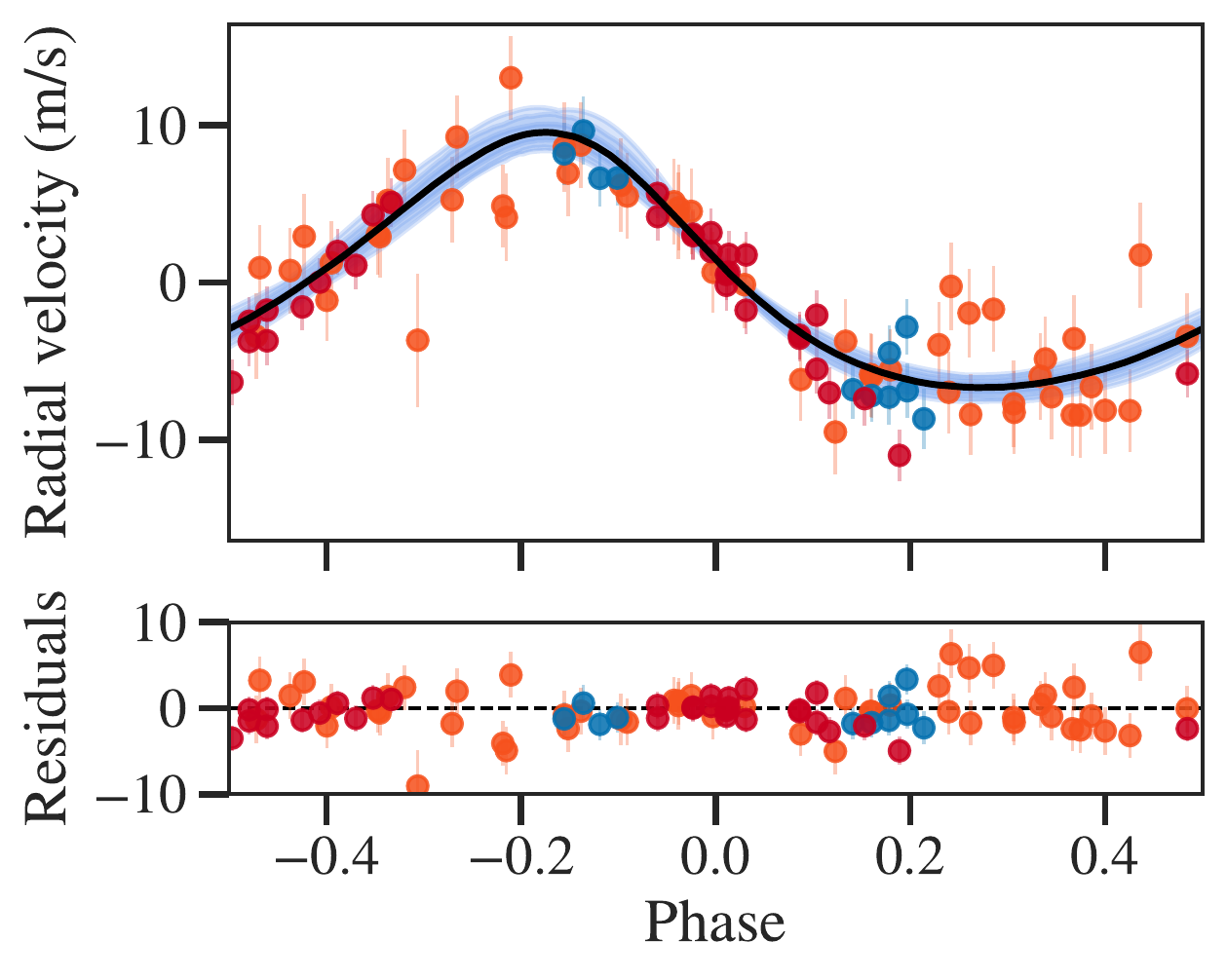}
\caption{Left: TESS lightcurve phased-folded to the period of 55 days. Solid line is the model for the transit. Bottom panel shows the residuals. Right: Phased-folded radial velocities from PFS1 (orange), HARPS (blue) and PFS2 (red) where the jitter has been added to the errobars. Solid black line represents the Keplerian model from the joint fit with \juliet. The orbital parameters for the system are listed in Table \ref{tab:planet}.}
\label{fig:jointfit}
\end{figure*}

For the photometry parameters we used the efficient sampling for the transit depth ($p$) and impact parameter ($b$) described in \citet{Espinoza2018efficient} that allows only physically plausible values in the ($b$,$p$) plane to be sampled via the  $r_{1}$ and $r_{2}$ coefficients according to the description of \citet{Kipping2013} for two parameter laws.  
As a result we obtained a planet mass of 42.44$^{+2.22}_{-2.08}\, M_{\oplus}$, consistent with a super-Neptune, with a radius of 3.89$^{+0.19}_{-0.20}\,R_{\oplus}$ that translates to a relatively high density of 3.98$^{+0.62}_{-0.64}$ \gcm~ for this planet. 
We note here we did not use GPs nor MA as in the radial velocity-only analysis, so the residuals shown in \ref{fig:jointfit} (right) are really the full residuals from a pure Keplerian model including instrumental jitter. 

\section{Additional Signals}
\begin{figure}
 \includegraphics[width=\columnwidth]{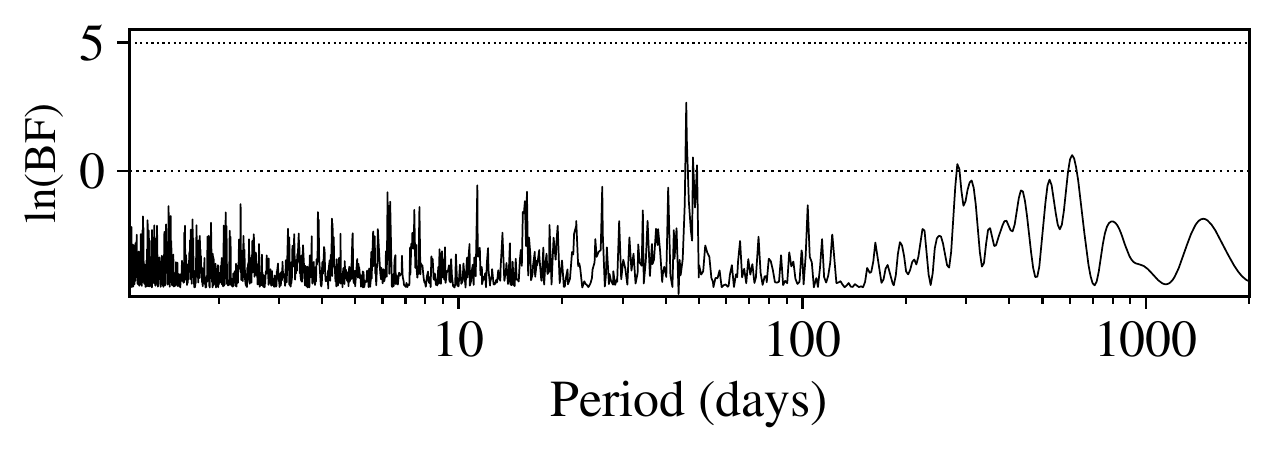}
\caption{Bayes Factor periodogram of the residuals for the 1-planet model from our joint fit with \texttt{juliet}. No statistically significant signals are seen after subtracting the 55-day period. There is a peak in the power spectrum around signal around 46 days, however it is below our detection threshold ln(BF)>5.}
\label{fig:residuals}
\end{figure}

{We searched for additional signals by analyzing the residuals from the 1-planet fit using same MA approach described in Section \ref{sec:bayes}.  Figure \ref{fig:residuals} shows the Bayes Factor Periodogram (\citealp[BFP;][]{Feng2017}) of the residual radial velocities for a 1-planet model. For this data, we do not find evidence for additional statistically significant signals present in the system after removing the 55-day planet signal.  However, we do see a periodic signal at $\sim$46 days in the residual BFP, but we cannot reach any conclusion at this moment as the signal is below the detection threshold of ln(BF)>5 to be considered as significant. It can be related to the activity of the star, based on what we see in the periodogram analysis of the stellar activity indicators where we see some hints of periodicities around 30-40 days. Additional spectroscopic data will help to confirm or rule out additional signals. }

\section{Discussion}

\begin{figure*}
\includegraphics[scale=0.8]{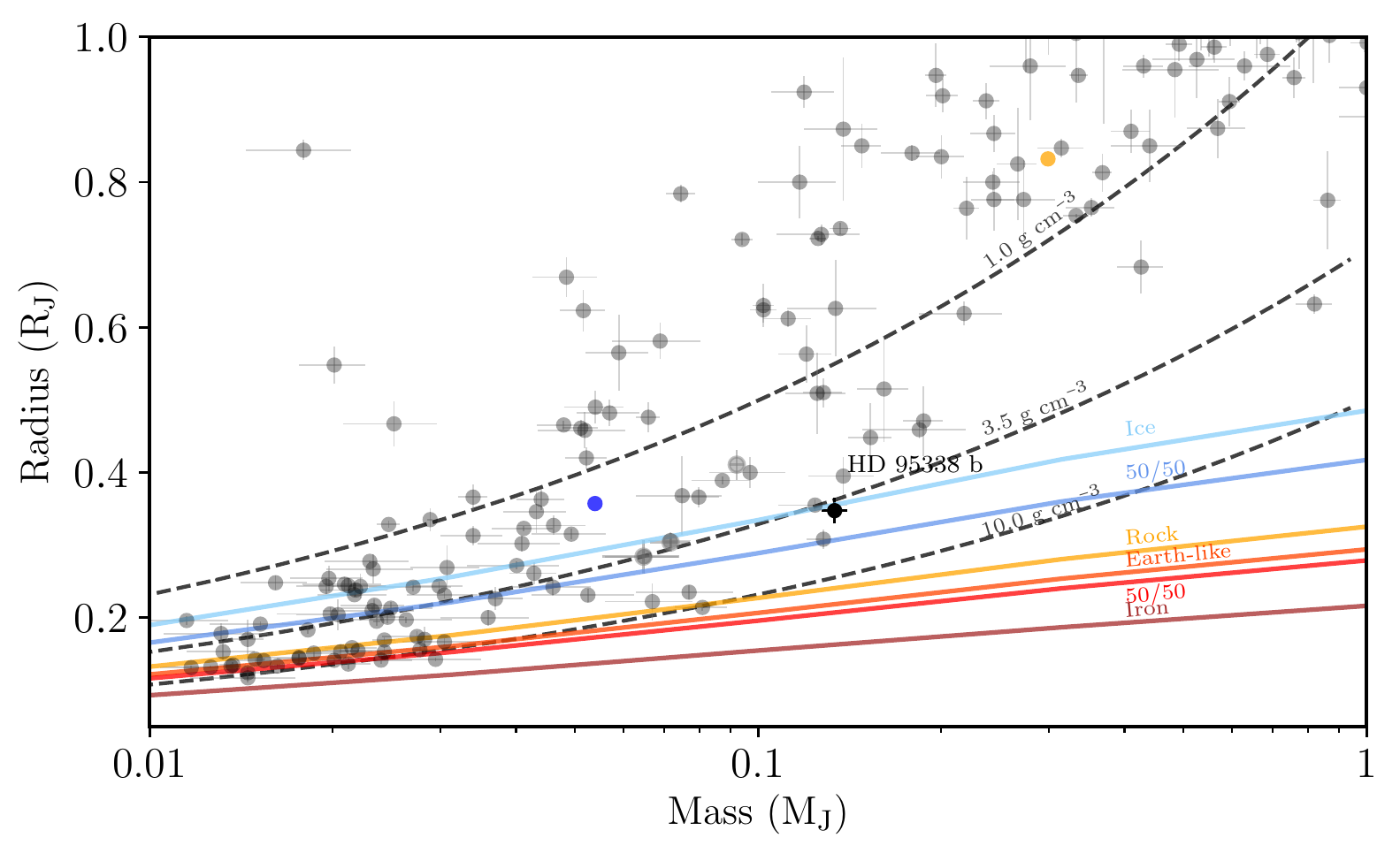}
\caption{Mass-radius diagram. Gray circles represent confirmed exoplanets from \texttt{TEPcat} \citep{Southworth2011} that have radius measurements with a precision of 20\% or better. Neptune (blue) and Saturn (yellow) are included for comparison. Three iso-density curves are represented by the grey dashed lines. Composition models are from \citet{Fortney2007}, and are shown by the coloured and labelled curves. The observed and derived parameters of HD 95338 b place this planet being consistent with an ice world (see text).}
\label{fig:mrplot}
\end{figure*}

To better understand the composition of HD 95338 b, we have constructed interior structure models matched to its observed mass, radius, and orbital parameters.  These models are explained in detail in \citet{Thorngren2016}; briefly, they solve the equations of hydrostatic equilibrium, conservation of mass, and the material equation of state to determine the radius of a well-mixed planet.  The equations of state (EOS) used were \citet{Chabrier2019} for H/He and a 50-50 ice-rock mixture from ANEOS \citep{Thompson1990} for the metals.  Giant planets gradually cool by radiating away the residual heat left over from their initial formation, which we regulated using the atmosphere models of \citet{Fortney2007} to evolve the planets through time.  Finally we used the Bayesian retrieval framework from \citet{Thorngren2019} to infer the bulk metallicities consistent with the planet parameters.  The planet is cool enough that no anomalous heating effect should be present. {The composition is consistent with that of ice (Figure \ref{fig:mrplot}), which is to say a mixture of ammonia, water, and methane without regard for the actual state of matter.  Indeed, the ices in this planet would be mostly supercritical fluids, with possibly plasma near the core, and maybe a small amount of gaseous water in the atmosphere.  The only solid material would be iron and rocks.}
 
 \begin{figure}
 \includegraphics[width=\columnwidth]{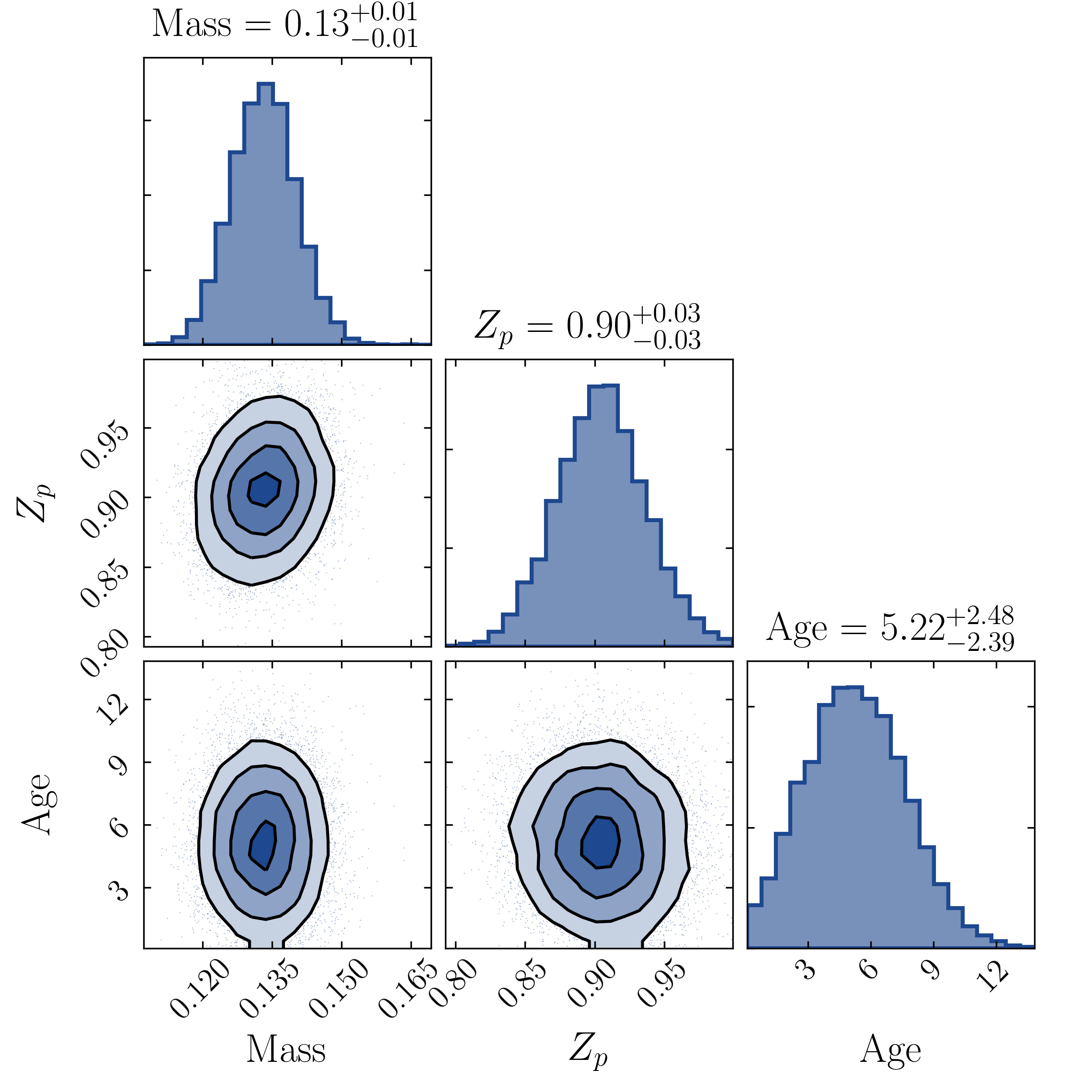}
 \caption{Corner plot showing the posteriors of heavy element content derived from the Bayesian retrieval framework described in  \citet{Thorngren2019}.}
 \label{fig:Zposteriors}
 \end{figure}

Our models show that to reproduce the planet’s high bulk density ($\rho_{\rm p}$=  3.98$^{+0.62}_{-0.64}$ \gcm), a metallicity of $Z$=0.90$\pm$0.03 was required (see Figure \ref{fig:Zposteriors}).  As such, it is among the most metal rich planets of this mass range, and raises questions about how the planet formation process can gather so much metals without also accreting more H/He. While extreme, this is not truly an outlier: other planets in this mass range are also found to have high metallicities (see \citealt{Thorngren2016}), including Kepler-413 b ($M_{\rm p}= 0.21 \, M_{\rm J}$, $Z \simeq 0.89$, \citealt{Kostov2014}) and K2-27 b ($M_{\rm p}=0.09\,M_{\rm J}$, $Z\simeq0.84$, \citealt{VanEylen2016}).  It could be that these highly metallic, and massive planets, were formed through collisions with other worlds after the proto-planetary disk had dispersed, stripping the planet of gas whilst enriching it with further metals. Indeed the results here imply that the heavy element enrichment for HD 95338~b is of order $\sim 38\, M_{\oplus}$. It is important to note that the radius measurement of this planet is sufficiently precise that modeling uncertainties are larger than statistical uncertainties.  These principally include uncertainties in the EOS, the interior structure of the planet (core-dominated vs well mixed), and the rock-ice ratio of the metals.  However, these uncertainties do not endanger the qualitative conclusion that the planet is extremely metal-rich, and changes would often lead to an even higher inferred $Z$.

\section{Conclusions}\label{sec:conc}
We present the discovery of a dense Neptune planet, that is currently the longest period planet known to transit a star brighter than $V=9$. Moreover it is the first single transit confirmed planet from the \tess~ mission.  It orbits the early-K star, HD 95338, and was originally detected using long-term radial velocity measurements carried out as part of the Magellan/PFS Exoplanet Survey. Additional radial velocity data from HARPS help to further constrain the period and orbital parameters of the candidate. \tess\, photometry shows a single transit observed in Sector 10. From our orbital parameters we estimated the transit time, $T_{c}=2458585.929\pm0.84$ and found it to be consistent within the errors with the observed transit by \tess, $T_{c,{TESS}}=2458585.279$, strongly suggesting both signals originate from the same source, and adding credibility to the reality of the planetary nature of the object. {After performing a joint model fit combining the radial velocities and the photometric measurements, we find the planet has a radius of $R_{\rm p}$=3.89$^{+0.19}_{-0.20}\,R_{\oplus}$  and a mass of $M_{\rm p}$=42.44$^{+2.22}_{-2.08}\, M_{\oplus}$, giving rise to an anomalously high density for this planet of $\rho_{\rm p}$=  3.98$^{+0.62}_{-0.64}$ \gcm. Planet structure models place HD 95338 b as being consistent with an ice world based on its mass and radius. From our Bayesian retrieval framework we estimated the heavy element content to be $Z=0.90\pm0.03$, which translates to $\sim38\, M_{\oplus}$. Such a high metallic value requires additional modeling efforts to explain and therefore
follow-up observations are crucial to arrive at a better understanding of the properties of the planet and also to further constrain models for how such a world could form in the first place. Moreover, the study of spin-orbit alignment of the planet with respect to the star via Rossiter-McLaughlin observations could provide some insights on the past history of the system such as interaction with companions and migration.}

\section*{Acknowledgements}
We thank N. Espinoza for useful discussion during the preparation of the manuscript. MRD acknowledges the support of CONICYT-PFCHA/Doctorado Nacional-21140646, Chile. JSJ acknowledges support by FONDECYT grant 1161218 and partial support by CATA-Basal (PB06, CONICYT). JV acknowledges support of CONICYT-PFCHA/Doctorado Nacional-21191829. D. D. acknowledges support from NASA through Caltech/JPL grant RSA-1006130 and through the TESS Guest Investigator Program Grant 80NSSC19K1727. Z.M.B. acknowledges funds from CONICYT/FONDECYT postdoctorado 3180405. This paper includes data collected by the \tess\, mission. Funding for the TESS mission is provided by the NASA Explorer Program.




\bibliographystyle{mnras}
\bibliography{references.bib} 

\bsp	
\label{lastpage}
\end{document}